\documentclass[journal]{IEEEtran} 
\ifCLASSINFOpdf
\else
\fi

\usepackage{cite}
\usepackage{stfloats}
\ifCLASSINFOpdf
   \usepackage[pdftex]{graphicx}
   \graphicspath{{../pdf/}{../jpeg/}}
   \DeclareGraphicsExtensions{.pdf,.jpeg,.png}
\else
   \usepackage[dvips]{graphicx}
   \graphicspath{{../eps/}}
   \DeclareGraphicsExtensions{.eps}
\fi

\usepackage[cmex10]{amsmath}
\usepackage{mathrsfs,color,amssymb}
\usepackage{amsfonts}

\usepackage{algorithm}
\usepackage{algorithmic}
\allowdisplaybreaks

\ifCLASSOPTIONcompsoc
\else
  \usepackage[caption=false,font=footnotesize]{subfig}
\fi

\usepackage{hyperref} 
\hypersetup{hidelinks}

\usepackage{graphicx}
\usepackage{epstopdf}

\newtheorem{theorem1}{\bf{Theorem}}
\newtheorem{lemma}{\bf{Lemma}}
\newtheorem{corollary}{\bf{Corollary}}
\newtheorem{defn}{\bf{Definition}}

\begin{document}

%

\title{\LARGE{Interleaved Training and Training-Based Transmission Design for Hybrid Massive Antenna Downlink}
}
%
%
%
\author{Cheng~Zhang,~\IEEEmembership{Student Member,~IEEE,}
        Yindi~Jing,~\IEEEmembership{Member,~IEEE,}
        Yongming~Huang,~\IEEEmembership{Senior Member,~IEEE,}
        Luxi~Yang,~\IEEEmembership{Member,~IEEE}
        \vspace{-5mm}
\thanks{
This work was supported in part by the National Natural Science Foundation of China under Grants 61720106003, in part by the Research Project of Jiangsu Province under Grant BE2015156, in part by the Scientific Research Foundation of Graduate School of Southeast University. Part of this work has been accepted by IEEE International Conference on Acoustics, Speech and Signal Processing (ICASSP) 2018 \cite{cheng_icassp_interleaved}. The guest editor coordinating the review of this manuscript and approving it for publication was Dr. Christos Masouros. (Corresponding author: Yongming Huang.)

C. Zhang, Y. Huang and L. Yang are with the National Mobile Communications Research Laboratory, Southeast University, Nanjing 210096, P. R. China (email:
{zhangcheng1988, huangym, lxyang}@seu.edu.cn).

Y. Jing is with the Department of Electrical and Computer Engineering, University of Alberta, Edmonton, Canada, T6G 1H9 (email: yindi@ualberta.ca).}
}

%
%

\markboth{Accepted by IEEE JSTSP. Copyright may be transferred without notice, after which this version may no longer be accessible. }
{Shell \MakeLowercase{\textit{et al.}}: Bare Demo of IEEEtran.cls for Journals}
%



\maketitle

\begin{abstract}
In this paper, we study the beam-based training design jointly with the transmission design for hybrid massive antenna single-user (SU) and multiple-user (MU) systems where outage probability is adopted as the performance measure. For SU systems, we propose an interleaved training design to concatenate the feedback and training procedures,  thus making the training length adaptive to the channel realization.
Exact analytical expressions are derived for the average training length and the outage probability of the proposed interleaved training. For MU systems, we propose a joint design for the beam-based interleaved training, beam assignment, and MU data transmissions.
Two solutions for the beam assignment are provided with different complexity-performance tradeoff.
Analytical results and simulations show that for both SU and MU systems, the proposed joint training and transmission designs achieve the same outage performance as the traditional full-training scheme but with significant saving in the training overhead.
\end{abstract}


%
\IEEEpeerreviewmaketitle

\vspace{-2mm}
 \section{Introduction}
Massive multiple-input-multiple-output (MIMO) is considered to be a promising technique to further increase the spectrum efficiency of wireless systems, thanks to the high spatial degrees-of-freedom it can provide
\cite{Hoydis_massive,Zhang_performance,Masouros_Towards}. However, the conventional full-digital implementation where one full radio-frequency (RF) chain is installed for each antenna tends to be impractical due to its high hardware costs and power consumptions \cite{Bogale_Hybrid}, especially for systems targeted at the millimeter wave (mmWave) band \cite{Gao_Reliable,Ming_mmw,Gao_Turbo}. Recently, enabled by the cost-effective and low-complexity phase shifters, a hybrid analog-digital structure has been applied for massive antenna systems which can effectively reduce the hardware complexity and costs via a combination of full-dimensional analog RF processing and low-dimensional baseband processing \cite{Sohrabi_hybrid,Alkhateeb_Channel}. It has been shown that this hybrid structure incurs slight performance loss compared with its full-digital counterpart \cite{Sohrabi_hybrid},
when perfect channel state information (CSI) is available.

One crucial practical issue for massive MIMO downlink is the acquisition of CSI at the base station (BS), especially when no channel reciprocity can be exploited, e.g., systems with frequency-division-duplexing (FDD) \cite{Zhang_Sum}. Due to the massive number of channel coefficients to be estimated \cite{Biguesh_Training}, traditional training and channel estimation schemes cause prohibitive training overhead. Thus, new channel estimation methods have been proposed for massive MIMO downlink by exploiting channel statistics \cite{Adhikary_Joint} or channel sparsity \cite{Rao_Distributed}.
While these apply to full-digital systems, they are less effective for the hybrid structure. Due to the limited number of RF chains and the phase-only control in the RF processing, it is more challenging to acquire the channel statistics or align with statistically dominant directions of the channels accurately and transmit sensing pilots with high quality \cite{Heath_overview}.

One popular method for the downlink training with hybrid massive antenna BS is to combine the codebook \cite{JJZ_codebook} based beam training with the traditional MIMO channel estimation \cite{Hur_millimeter,Wang_Beam,Alkhateeb_Channel}. In this method, a finite codebook is used which contains all possible analog precoders, called beams. Then the channel estimation problem is transformed to the estimation of the beam-domain effective channels. This can reduce the dimension of the channel estimation problem if the number of desired beams is limited. The remaining difficulty lies in finding the desired beams or analog precoders with affordable training overhead.

Several typical beam-based training schemes have been proposed for hybrid massive MIMO downlink.
For systems with single RF chain at the BS, which can serve single user (SU) only, one straightforward method is to exhaustively train all possible beams in the codebook, then to find the best beam for transmission. Another typical method is based on hierarchical search \cite{Hur_millimeter, Alkhateeb_Channel}, where all possible wide beams are trained first and the best is selected. Within this selected best wide beam, the beams with narrower beamwidth are trained and the best is chosen. The training procedure is repeated until the optimal beam with acceptable beamwidth is found.
Generally speaking, the hierarchical search scheme has lower training overhead than the scheme with exhaustive search. However, since hierarchical search uses wide beams first, its beam alignment quality is very sensitive to the pre-beamforming signal-to-noise-ratio (SNR) \cite{Liu_millimeter}. Meanwhile, its advantage of lower training overhead diminishes as the number of channel paths increases or when applied to multiple-user (MU) systems \cite{Alkhateeb_Channel, Alkhateeb_compressed}. 

Beam training procedure for hybrid massive antenna systems with multiple RF chains is analogous to that of the single RF chain case.
The users generally feed back channels of multiple beams \cite{Lee_Impact}, and then the BS selects the best beam combination, and constructs the corresponding analog precoder and baseband precoder for data transmission. The work in \cite{Amadori_Low} studied the beam selection problems with respect to the capacity and signal-to-interference-plus-noise-ratio (SINR) maximization.
In \cite{Ghadikolaei_Beam}, the tradeoff of transmission beamwidth and training overhead was studied, where the beamwidth selection and pair scheduling were jointly designed to maximize network throughput. In \cite{Lee_Impact}, partial beam-based training was proposed where only a subset of the beams are trained and the sum-rate loss compared with full training was studied. Another scheme proposed in literature is based on tabu search \cite{Gao_Turbo} where several initial beams are first chosen and the training procedure is only executed within the neighbors obtained by  changing the beam of only one RF chain and fixing the others. The training stops when a local optimal beam combination is found in terms of mutual information.


For existing beam-based training and corresponding transmission schemes, the basic idea is to obtain the complete effective CSI for the full or selected partial beam codebook. Based on the obtained effective CSI values, data transmission designs are then proposed. In such schemes, the training design and data transmission design are decoupled. For satisfactory performance, the size of the training beam codebook needs to increase linearly with the BS antenna number, leading to heavy training overhead for massive MIMO systems. The decoupled nature of existing schemes also imposes limitations on the tradeoff between training overhead and performance. Further, only the throughput or diversity gain has been considered in existing work. In \cite{Koyuncu_interleaving}, an interleaved training scheme was proposed for the downlink of SU full-digital massive antenna systems with independent and identically distributed (i.i.d.) channels.  In this scheme, the BS trains its channels sequentially and the estimated CSI or indicator is feed back immediately after each training step. With each feedback,
the BS decides to train another channel or terminate the training process based on whether an outage occurs. The scheme was shown to achieve significant reduction in training overhead with the same outage performance compared to traditional schemes.

By considering the aforementioned limitations of existing training schemes and exploiting the interleaved training idea, in this paper, we study the beam-based training design jointly with the data transmission design for hybrid massive antenna systems with single user and multiple users.
The outage probability is adopted as the performance measure. We consider interleaved training designs that are dynamic and adaptive, in the sense that the length of the training interval depends on the channel realization and the termination of the training process depends on previous training results, to achieve favorable tradeoff between the outage performance and the training overhead. 
Compared with the work in \cite{Koyuncu_interleaving}, our work is different in the following aspects. 
First, \cite{Koyuncu_interleaving} is on full-digital massive MIMO systems with  i.i.d.~channels. Whereas, in this work, we consider hybrid massive antenna systems and a more general channel model that incorporates channel correlation and limited scattering. Our work uses beam-based training while in \cite{Koyuncu_interleaving} the training is conducted for each antenna sequentially. 
Second, given the above differences in system, channel, and transceiver models, the theoretical derivations and analytical results in our work are largely different. The numbers of resolvable channel paths and RF chains are two new parameters in our work. They both make the derivations on the average training length and the outage probability considerably more complicated compared to the case in \cite{Koyuncu_interleaving}. Our work also provides more comprehensive analytical performance results along with abundant new insights.  
Third, this work considers both SU and MU systems. Especially for MU systems, new design issues appear, such as the beam assignment and the local CSI at each user. Although the same basic interleaved and adaptive training idea is
used, the implementation of this idea for MU systems is far from immediate applications and
the algorithm has large difference to the one for the SU case. 
The distinct contributions of this work are summarized as follows.
\begin{itemize}
\item For SU massive antenna systems with arbitrary number of RF chains, we propose a beam-based interleaved training scheme and the corresponding joint data transmission design. The average training length and the outage probability of the proposed scheme are studied, where exact analytical expressions are derived.
\item For MU massive antenna systems with arbitrary number of RF chains, we propose a joint beam-based interleaved training and data transmission design. 
Two beam assignment solutions, i.e., exhaustive search and max-min assignment, are proposed with different complexity-performance tradeoff. Compared to exhaustive search, the low-complexity max-min method induces negligible increment in the average training length and small degradation in the outage performance. Due to its advantage in complexity, the max-min method is more desirable for the proposed MU scheme. 
\item Analytical results and simulations show that for both SU and MU systems, the proposed training and joint transmission designs achieve the same outage probability as the traditional full-training scheme but with significant saving in the training overhead. 
\item Based on the analytical results and simulations, useful insights are obtained on the performance of several special but typical scenarios, e.g., small channel angle spread (AS) or limited scattering, and also on the effect of important system parameters, e.g., the BS antenna number, the RF chain number, the channel path number or AS, and the rate requirement, on the average training length and the outage performance.
\end{itemize}

Specifically, for the average training length of the proposed  SU scheme, our analysis reveals the following. 1) For channels with limited scattering, the average training length is no longer a constant as in the case of i.i.d.~channels, but linearly increases with respect to the BS antenna number. However, it decreases linearly with increasing channel paths. Meanwhile, fewer RF chains or higher rate requirement has negligible effect on the average training length. 
2) For channels with non-negligible AS, the average training length is a constant dependent on the AS, the RF chain number, and the rate requirement. The constant increases for higher rate requirement while the increasing slope decreases with more RF chains. Moreover, smaller AS (larger channel correlation) reduces the increasing speed of the average training length with higher rate requirement. For the outage probability of the proposed scheme, the following major new insights are achieved. 1) For channels with limited scattering and single RF chain, the outage probability scales as the reciprocal of the BS antenna number to the power of the channel path number.
2) For channels with non-negligible AS and single RF chain, the outage probability decreases exponentially with respect to the BS antenna number. 3) More RF chains can further decrease the outage probability for both kinds of channels. 

\textit{Notation:} In this paper, bold upper case letters and bold lower case letters are used to denote matrices and vectors, respectively. For a matrix $\bf A$, its conjugate transpose, transpose, and trace are denoted by ${\bf A}^H$, ${\bf A}^T$ and ${\rm tr}\{{\bf A}\}$, respectively. For a vector $\bf a$, its conjugate counterpart is ${\bf a}^*$. ${\rm E}[\cdot]$ is the mean operator and ${\rm Pr}[\cdot]$ indicates the probability. The notation ${a} = \mathcal{O}\left( {{b}} \right)$ means that ${a}$ and ${b}$ have the same scaling with respect to a parameter given in the context.
$\|\bf a\|$ denotes the 2-norm of $\bf a$ and $\|{\bf A}\|_F$ denotes the Frobenius norm of $\bf A$.
$\Upsilon \left( {s,x} \right) = \int_0^x {{t^{s - 1}}{e^{ - t}}} dt$ is the lower incomplete gamma function and $\Gamma\left( {s,x} \right)=\int_x^\infty {{t^{s - 1}}{e^{ - t}}} dt$ is the upper incomplete gamma function. 
${\mathcal X}^2(k)$ denotes the chi-squared distribution with $k$ degrees of freedom.
$\mathcal{CN}({\bf 0},{\bf \Sigma})$ denotes the circularly symmetric complex Gaussian distribution where the mean vector is $\bf 0$  and the covariance matrix is ${\bf \Sigma}$.

\section{System Model and Problem Statement}
\subsection{System Model}\label{secB_1}
Consider the downlink of a single-cell\footnote{This work only considers single-cell systems without inter-cell interference. The main reason for the simplification is to focus on  interleaved training design and fundamental performance behavior. The multi-cell case is left for future work.} massive antenna system with hybrid beamforming structure.
The BS employs $N_t\gg1$ antennas with $N_{RF}\in[1,N_t)$ RF chains and serves $U$ single-antenna users.
 Since for effective communications, each user requires a distinct beam, $U\le N_{RF}$ is assumed. Let ${\bf H}=[{\bf h}_1,...,{\bf h}_U]\in \mathbb{C}^{N_t\times U}$ be  the downlink channel matrix.

\subsubsection{Channel Model}\label{channelmodel}
We consider the typical uniform array, e.g., uniform linear array or uniform planar array, at the BS and high dimension.
The beamspace representation of the channel matrix becomes a natural choice \cite{Heath_overview}, \cite{Brady_Beamspace} where the antenna space and beamspace are related through a spatial discrete Fourier transform (DFT). Denote the DFT matrix as ${\bf D} \in \mathbb{C}^{N_t\times N_t}$ where the $i$th column is ${\bf d}_{i}=[1,e^{-j2\pi(i-1)/N_t},...,e^{-j2\pi(i-1)(N_t-1)/N_t}]^T,\forall i$. Assume that there are $L\in[1,N_t]$ distinguishable scatterers or paths \cite{Heath_overview} in User $u$'s channel $\forall u$, and define the set of their direction indices as  ${\mathcal I}_u=\{I_{u,1},...,I_{u,L}\}$. The channel vector of User $u$ can be written as
\begin{equation}\label{ch1}
{\bf{h}}_u={\bf D}{\bf \bar h}_u=[{\bf d}_{1},...,{\bf d}_{N_t}][{ \bar h}_{u,1},...,{ \bar h}_{u,N_t}]^T,
\end{equation}
where ${ \bar h}_{u,i}\sim \mathcal{CN}(0,1/L)$ for $i\in {\mathcal I}_u$ and ${ \bar h}_{u,i}=0$ for $i\notin {\mathcal I}_u$.
This channel model can be understood as an asymptotic approximation of the geometric channel model in \cite{Tse_Funda} which has been widely used for the mmWave band \cite{Alkhateeb_Channel,Alkhateeb_Limited}
with discretized angle distribution of the channel paths.

Specifically, we assume that different users have independent path directions and gains, and the $L$-combination $(I_{u,1},\cdots,I_{u,L})$ follows discrete uniform distribution with each element on $[1,N_t]$.
One justification is given as follows. Generally, for the geometric channel model, the angles of different paths are independent following uniform distribution \cite{Alkhateeb_Channel} and no two paths' continuous angles are the same. The beamspace representation equivalently divides the angle space into $N_t$ uniform sections\cite{Heath_overview}. When $N_t$ is large enough, no two paths are in the same section. As the variances of ${\bar h}_{u,i}, i\hspace{-1mm}\in\hspace{-1mm} {\mathcal I}_u$ are set to be the same, the average power difference among different paths is not embodied in this channel model. Further, it is assumed that all users have the same $L$.
When $L=N_t$, our channel model becomes the i.i.d.~one \cite{Koyuncu_interleaving}. When $L=1$, it reduces to the single-path one \cite{Lee_Impact}.

While generally speaking, the number of distinguishable channel paths $L$ is arbitrary in our work, two typical scenarios are of special interest, corresponding to  different scaling with respect to $N_t$. The first typical scaling for $L$ is that it is a constant with respect to $N_t$, i.e., $L=\mathcal{O}(1)$. This corresponds to
channels with extremely small AS where having more antennas does not result in more distinguishable paths. One application is the outdoor environment with a few dominant clusters \cite{Heath_overview}.
Another typical scaling is when $L$ linearly increases with $N_t$, i.e., $L=cN_t$ with $c\in(0,1]$ being a constant. It corresponds to channels with non-negligible AS where $c$ is the value of the AS. Since the spatial resolution increases linearly with $N_t$, it is reasonable to assume that the number of distinguishable path increases linearly with $N_t$. One application is the indoor environment with a large amount of reflections \cite{Heath_overview}.
Similarly, two typical scalings for $N_{RF}$ are $N_{RF}=\mathcal{O}(1)$ and $N_{RF}={\bar c}N_t$ with $\bar c\in(0,1]$ being a constant. The former case is more interesting from the perspective of low hardware costs.

\subsubsection{Hybrid Precoding and Outage Probability}\label{HP-OP}
The hybrid structure at the BS calls for an analog RF precoding followed with a baseband precoding.

The analog precoder ${\bf F}_{RF} \in \mathbb{C}^{N_t\times L_s} $ is realized by phase shifters for low hardware complexity, where $L_s$ is the number of beams used for the transmission and $L_s\le N_{RF}$. All elements of ${\bf F}_{RF}$ have the same constant norm. Without loss of generality, we assume $\|{[{\bf F}_{RF}]}_{i,j}\|^2=1/N_t, \forall i,j$. The codebook-based beamforming scheme is used in this work, where columns of ${\bf F}_{RF}$ are chosen from a codebook of vectors $\mathcal{F}_{RF}$ \cite{Alkhateeb_Channel}, \cite{Hur_millimeter}. Naturally, with the channel model in (\ref{ch1}), the DFT codebook is used \cite{He_Codebook}, where $\mathcal{F}_{RF}=\{{\bf d}^*_1/\sqrt{N_t},\cdots, {\bf d}^*_{N_t}/\sqrt{N_t}\}$.
Each element in the codebook is also called a beam and there are $N_t$ beams in total. With a given analog beamforming matrix ${\bf F}_{RF}$, the effective channel matrix for the baseband is ${\bf H}^{T}{\bf F}_{RF}$. More specifically, ${\bf h}_u^T{\bf d}^*_i/\sqrt{N_t}=\sqrt{N_t}{\bar h}_{u,i}$ is the effective channel of User $u$ on Beam $i$. If ${\bar h}_{u,i}\ne 0$, Beam $i$ is a non-zero beam for User $u$.

The next to discuss is the baseband precoding and the outage probability. In what follows, we consider the SU case and the MU case separately due to their fundamental difference.

For the SU case (i.e., $U=1$) where the BS chooses $L_s$ beams for analog precoding,
the transmitted signal vector is $\sqrt{P}{\bf F}_{RF}{\bf f}_{BB}{s}$ and the transceiver equation can be written as
\begin{equation}
{y}=\sqrt{P}{\bf h}^T{\bf F}_{RF}{\bf f}_{BB}{s} + { n}=\sqrt{P}{\bf \bar h}^T{\bf D}{\bf F}_{RF}{\bf f}_{BB}{s} + { n},
\end{equation}
where $P$ can be shown to be the short-term total transmit power, $\bf h$ is the channel vector from the BS to the user, ${\bf f}_{BB}$ is the baseband beamformer, $s$ denotes the data symbol with unit power, and $n$ is the additive noise following $\mathcal{CN}(0,1)$. For a fixed $\bf h$, with perfect effective channel vector, i.e., ${\bf \bar h}^T{\bf D}{\bf F}_{RF}$, at the user side, the channel capacity is $\log_2(1+P\|{\bf \bar h}^T{\bf D}{\bf F}_{RF}{\bf f}_{BB}\|^2)$ bps/Hz. For a given  transmission rate $R_{th}$, an outage event occurs if 
\begin{equation}
\|{\bf \bar h}^T{\bf D}{\bf F}_{RF}{\bf f}_{BB}\|^2\le\alpha\triangleq \frac{2^{R_{th}}-1}{P},
\end{equation}where $\alpha$ is called the \textit{target normalized received SNR}. Thus, for random $\bf h$, the outage probability is
\begin{equation}\label{definition_OP_SU}
{\rm out}({\bf F}_{RF}, {\bf f}_{BB})={\rm Pr}(\|{\bf \bar h}^T{\bf D}{\bf F}_{RF}{\bf f}_{BB}\|^2\le\alpha).
\end{equation}
If further the effective channel vector is perfectly known at the BS, ${\bf f}_{BB}$ can be designed to match the effective channel vector, i.e., ${\bf f}_{BB}={({\bf \bar h}^T{\bf D}{\bf F}_{RF})^H}/\|{\bf \bar h}^T{\bf D}{\bf F}_{RF}\|$, which is optimal in the sense of minimizing the outage probability. In this case, the outage probability becomes ${\rm Pr}(\|{\bf \bar h}^T{\bf D}{\bf F}_{RF}\|^2\le \alpha)$ which is dependent on  ${\bf F}_{RF}$.

If more than $L_s$ non-zero beams are available, beam selection is needed. Define the set of indices of the available non-zero beams as $\mathcal{A}=\{a_1,...,a_j\}$. The optimal beam-selection is to find the strongest $L_s$ ones within the set. By ordering the magnitudes of the effective channels as $\|{\bar h}_{s_1}\|\hspace{-0.5mm}\ge\hspace{-0.5mm} \|{\bar h}_{s_2}\|\hspace{-0.5mm}\ge\hspace{-0.5mm}...\hspace{-0.5mm}\ge\hspace{-0.5mm}\|{\bar h}_{s_{L_s}}\|\hspace{-0.5mm}\ge\hspace{-0.5mm}\cdots\hspace{-0.5mm}\ge\hspace{-0.5mm}\|{\bar h}_{s_{j}}\|$. The set of indices of the selected beams is $\mathcal{S}=\{s_1,...,s_{L_s}\}$. Thus the beamforming matrices are:
\begin{eqnarray}
&&{\bf F}_{RF}={\left[\frac{{{\bf d}}^*_{s_1}}{\sqrt{N_t}},...,\frac{{{\bf d}}^*_{s_{L_s}}}{\sqrt{N_t}}\right]}, \nonumber \\
&&{\bf f}_{BB}=\frac{[{\bar h}_{s_1}, ... , {\bar h}_{s_{L_s}}]^{H}}{{\sqrt{\|{\bar h}_{s_1}\|^2+...+\|{\bar h}_{s_{L_s}}\|^2}}}.
\label{F_rf_bb}
\end{eqnarray}
The outage probability reduces to ${\rm Pr}\left(\sum_{i=1}^{L_s}\|{\bar h}_{s_{i}}\|^2\le\alpha/N_t\right)$.

For the MU case, we assume that the BS uses $U$ out of the $N_{RF}$ RF chains to serve the $U$ users, i.e., $L_s=U$. The received signal vector at the users can be presented as
\begin{equation}\label{eq1}
{\bf y}=\sqrt{P}{\bf H}^{T}{\bf F}_{RF}{\bf F}_{BB}{\bf s} + {\bf n},
\end{equation}
where ${\bf s}\sim \mathbb{C}^{U \times 1}$ contains the $U$ independent data symbols satisfying ${\rm E}[{\bf s}{\bf s}^H]=(1/U){\bf I}_{U}$;  and ${\bf n} \sim \mathcal{CN}({\bf 0}, {\bf I})$ is the additive noise vector. For the shot-term power normalization at the BS, we set
${\rm tr}\{{\bf F}_{RF}{\bf F}_{BB}{\bf F}_{BB}^H{\bf F}_{RF}^H\}=U$.

Without loss of generality, assume that Beams $n_1,\cdots,n_U$ are selected to serve Users $1,\cdots, U$ respectively. Thus ${\bf F}_{RF}=[{ {\bf d}}^*_{n_1}/\sqrt{N_t},...,{{\bf d}}^*_{n_U}/\sqrt{N_t}]$. The effective channel matrix ${\bf \hat{H}}={{\bf H}^{T}}{\bf F}_{RF}$ is therefore
\begin{eqnarray}\label{effective channel}
\hspace{-0.3cm}{\bf \hat{H}}&\hspace{-0.35cm}=\hspace{-0.35cm}&{[{\bf \bar h}_1,...,{\bf \bar h}_U]^{T}}{\bf D}{\bf F}_{RF}\hspace{-0.1cm}=\hspace{-0.1cm}\sqrt{N_t}\left[ {\begin{array}{*{20}{c}}
{{{{{\bar h}}}_{1,{n_1}}}}& \cdots &{{{{{\bar h}}}_{1,{n_U}}}}\\
 \vdots & \ddots & \vdots \\
{{{{{\bar h}}}_{U,{n_1}}}}& \cdots &{{{{{\bar h}}}_{U,{n_U}}}}
\end{array}} \right].
\end{eqnarray}
One of the most widely used baseband precodings is the zero-forcing (ZF) scheme \cite{Amadori_Low}, \cite{Alkhateeb_Limited}: ${\bf F}_{BB}=\lambda{\bf \hat{H}}^{H}({\bf \hat{H}}{\bf \hat{H}}^H)^{-1}$, where
\begin{equation}
\lambda=\sqrt{{U}/\|{{\bf F}_{RF}{\bf \hat{H}}^{H}({\bf \hat{H}}{\bf \hat{H}}^H)^{-1}}\|_F^2}.
\label{lambda}
\end{equation}
With ZF baseband precoding, the user-interference is cancelled and the received SNRs of all users are the same, which are
\begin{equation}\label{mu_SINR}
{\rm SNR}_{\rm MU}=(P/U)\lambda^2.
\end{equation}

For a given target per-user transmission rate $R_{th}$, an outage event occurs for User $u$ if
${\rm SNR}_{\rm MU}\le (P/U)\bar \alpha,$
where 
\begin{equation}
\bar \alpha\triangleq \frac{2^{R_{th}}-1}{P}U
\end{equation}is the \textit{target normalized per-user received SNR}. The outage probability for the system with $U$ users is thus
\begin{equation}\label{eq_7}
{\rm out}({\bf F}_{RF}, {\bf F}_{BB})={\rm Pr}\left(\lambda^2\le \bar \alpha\right).
\end{equation}

\subsection{Beam-Based Training and Existing Schemes}
To implement hybrid precoding, including both the beam selection/assignment and the baseband matching/ZF, CSI is needed at the BS, thus downlink training and CSI feedback must be conducted. Instead of all entries in ${\bf H}$, for the hybrid massive MIMO system under codebook-based beamforming, the BS only needs values of the effective channels $\sqrt{N_t}{\bar h}_{u,i}$'s. Thus beam-based training is a more economical choice than traditional MIMO training \cite{Biguesh_Training}. In what follows, existing beam-based training schemes are briefly reviewed.

For SU systems, the typical beam-based training scheme operates as follows. For each channel realization, the BS sequentially transmits along the $N_t$ beams for the user to estimate the corresponding effective channels. The effective channel values are then sent back to the BS. Another type of beam-based training scheme uses hierarchical search \cite{Alkhateeb_Channel}, which generally has smaller training length. But this advantage diminishes as the path number $L$ increases or the pre-beamforming SNR decreases, along with performance degradation \cite{Liu_millimeter}.  
Specifically, for SU systems with $N_t$ BS antennas, $N_{RF}$ RF chains, and  $L$ channel paths,  the training length of the hierarchical scheme is $T_{\rm HT-SU}=ML^2\log_M(N_t/L)$, where  $M\ge 2$ is the partition number for the beamwidth, e.g., $M=2$ means bisection \cite[Section V-B-1]{Alkhateeb_Channel}. Note that the training length is independent of $N_{RF}$ since the ideal hierarchical codebook is assumed. Meanwhile, the mechanism of the hierarchical scheme needs $L\le N_t/M<N_t/e^{0.5}=N_t/1.65$ for plausible results. It can be shown that the training length increases with $L$ for the aforementioned range of $L$.
As for the tabu search based training \cite{Gao_Turbo},  it is sensitive to the initialization step and stops when a local optimal beam combination is found. Thus it has worse outage performance.

For MU systems, beam-based training has been studied in \cite{Lee_Impact}, \cite{Alkhateeb_Limited}, \cite{He_Codebook} with similar procedure to the SU case reviewed above since all users can conduct channel estimation at the same time when the BS sends a pilot along one beam. But in \cite{Lee_Impact}, a more general scheme was proposed where only $L_t$ out of $N_t$ beams are selected for training. The value of $L_t$ can be used to leverage the tradeoff between the training overhead and the performance. Larger $L_t$ means longer training period, less time for data transmission, and better transmission quality; while smaller $L_t$ leads to the opposite. For the special case of $L_t=N_t$, the scheme becomes the full training case in \cite{He_Codebook}.

It should be noted that while the training procedure for the MU case is similar to that for the SU case, the effective CSI feedback and the beam assignment at the BS are different. Specifically, for an MU system with $L_t$ beams being trained, assume that there are $j_u$ non-zero beams among the trained beams for User $u, \forall u$. Then User $u$ feeds back to the BS the effective channels of the $j_u$ non-zero beams along with their indices. And the BS finds the beam assignment based on the CSI feedback. The work in \cite{Amadori_Low} considered the magnitude of the path and the SINR, while the sum-rate maximization was used in \cite{Lee_Impact, Alkhateeb_Limited}.

{\color{blue}}


\subsection{Motivations of This Work}
This work is on beam-based training design for SU and MU massive MIMO systems with hybrid structure. The object is to propose beam-based training schemes and corresponding SU and MU transmission schemes with reduced training length, without sacrificing the performance compared with the full training case. The training length, or the number of symbol transmissions required in the training period, is a crucial measure for training quality since it affects both the available data transmission time and the beamforming/precoding gain during transmission. In existing beam-based training schemes, the training length is fixed regardless of the channel realization and further the effective CSI feedback is separated from the training procedure. Thus we refer such designs as \textit{non-interleaved training (NIT)}. The combination of the NIT and data transmission for SU and MU systems are referred to as  NIT-SU and NIT-MU transmission schemes, respectively. For our object, \textit{interleaved training} idea is used, where for each channel realization, the training length is adaptive to the channel realization. Further, the effective CSI or indicator feedback is concatenated with the pilot transmissions to monitor the training status and guide the action of the next symbol period. Naturally, for interleaved schemes, the training and the data transmission need to be designed jointly.

In addition, in this work, outage probability is used as the major performance measure, which is different to most existing work where the sum-rate \cite{Alkhateeb_Limited} and SINR \cite{Amadori_Low} are used. While outage probability is a useful practical quality-of-service measure for wireless systems, its adoption in massive MIMO is very limited \cite{Feng_Interference} due to the high challenge in the analysis. With outage probability as the performance measure and the aim of reducing training length, in the proposed interleaved schemes, the basic idea is to stop training when the  obtained effective CSI is enough to support the targeted rate to avoid an outage. Other than training designs, quantitative analysis will be conducted on the outage performance of the proposed schemes for useful insights in hybrid massive antenna system design.

\section{Interleaved Training for SU Systems and Performance Analysis}
This section is on the SU system, where interleaved beam-based training and the corresponding SU transmission are proposed.  Further, outage probability performance is analyzed as well as the average training length of the proposed scheme.

\subsection{Proposed Interleaved Training and SU Transmission}

Recall that the object of interleaved training is to save training time while still having the best outage probability performance. Thus, instead of training all beams and finding the best combination as in NIT, the training should stop right after enough beams have been trained to avoid outage.
Since the set of $L$ non-zero beams for the user $\mathcal{I}$ is random with uniform distribution on the set $\{1,\cdots,N_t\}$, and the channel coefficients along the non-zero beams are i.i.d., the priorities of the training for all beams are the same. Therefore, the natural order is used for beam training, i.e., the BS trains from the first beam to the $N_t$-th beam sequentially. The training stops when the outage can be avoided based on the already trained beams or no  more beam is available for training.

Let $\mathcal{B}$ contain the indices of the non-zero beams that have been trained. Let $L_B\triangleq\min(N_{RF},|\mathcal{B}|)$ be the maximum number of non-zero beams that can be used for data transmissions given the number of RF chains and the number of known non-zero beams. Let $\mathcal{S}$ contain the indices of the $L_B$ known non-zero beams with the largest norms.
The proposed interleaved training and the corresponding SU transmission scheme are shown in Algorithm \ref{ISTS}.
\begin{algorithm}[htb]
\caption{Proposed Interleaved Training and Corresponding SU Transmission (IT-SU) scheme}
\label{ISTS}
\begin{algorithmic}[1]
\STATE $\mathcal{B}=\emptyset$;
\FOR{$i=1,...,N_t$}
\STATE The BS trains the $i$th beam; The user estimates $\bar{h}_{i}$;
\STATE If $\|{\bar h}_{i}\|>0$, $\mathcal{B}=\mathcal{B}\cup \{i\}$ and the user finds $\mathcal{S}$, which contains the indices of the $L_B$ non-zero beams with the largest norms, then calculates $\sum_{l\in \mathcal{S}}\|{\bar h}_{l}\|^2$;
\IF {$\|{\bar h}_{i}\|=0$ or $\sum_{l\in \mathcal{S}}\|{\bar h}_{l}\|^2 \le \alpha/N_t$}
\STATE The user feeds back ``0"; \textit{Continue};
\ELSE
\STATE The user feeds back ${\bar h}_{l}$, for all $l\in \mathcal{S}$ along with their indices;
\STATE The BS constructs ${\bf F}_{RF}$ and ${\bf f}_{BB}$ as in \eqref{F_rf_bb} and conducts data transmission; \textit{Break;}
\ENDIF
\ENDFOR
\end{algorithmic}
\end{algorithm}

In the proposed scheme, at the $i$th training interval where $i\le N_t$, the BS sends a pilot for the user to estimate the $i$th beam value: ${\bar h}_i$ (the scalar $\sqrt{N_t}$ is omitted for brief notation)\footnote{In this work, the channel estimation is assumed to be error-free. For massive antenna systems, the post-beamforming SNR during the training phase is usually high, leading to small channel estimation error. Meanwhile,
this simplification has little effect on the structure of the proposed scheme, although it may cause degradation in the outage performance.}. If it is a non-zero beam (i.e., $\|{\bar h}_{i}\|>0$), the user compares the received SNR provided by the $L_B$ strongest beams known from the finished $i$ training intervals with the target normalized received SNR $\alpha$ to see if an outage event will happen given the obtained CSI.
If $\|{\bar h}_{i}\|=0$ or $\sum_{l\in \mathcal{S}}\|{\bar h}_{l}\|^2 \le \alpha/N_t$ and $i<N_t$, the already trained beams cannot provide a beam combination to avoid outage. Thus the user feeds back the indicator ``0" to request the BS to continue training the next beam. For the special case of $i=N_t$, all beams have been trained and an outage is unavoidable with any beam combination. If $\sum_{l\in \mathcal{S}}\|{\bar h}_{l}\|^2 > \alpha/N_t$, enough beams have been trained to avoid outage. Thus the user feeds back the $L_B$ non-zero effective channels ${\bar h}_{l}, l\in \mathcal{S}$ along with their indices, and the BS aligns the $L_B$ beams with ${\bf F}_{RF}$ and matches the effective channel vector with ${\bf f}_{BB}$ as in \eqref{F_rf_bb} to conduct data transmission. Since the training and feedback are interleaved in the proposed scheme, we name it interleaved training based SU transmission (IT-SU) scheme.

\subsection{Average Training Length Analysis}
\label{sec-SU-time}
This subsection studies the average training length of the IT-SU scheme. Since for different channel realizations, the number of beams being trained in our proposed IT-SU scheme can vary due to the randomness in the path profile and path gains, we study the average training length measured in the number of training intervals per channel realization where the average is over channel distribution.

Before showing the analytical result, we first discuss the effect of $N_{RF}$ on the average training length. Intuitively, for any given channel realization and at any step of the training process, larger $N_{RF}$ means that the same or more beam combinations are available based on the already trained beams. Thus the same or larger received SNR can be provided, which results in the same or an earlier termination of the training period. Therefore, with other parameters fixed, the average training length is a non-increasing function of $N_{RF}$, i.e., larger $N_{RF}$ helps reduce the average training length.
On the other hand, since there are at most $L$ non-zero paths in the channel for each channel realization, having a larger $N_{RF}$ than $L$ cannot provide better beam combination for any given already trained beams compared with that of when $N_{RF}=L$. Therefore, with other parameters fixed, the average training length of the IT-SU scheme is a non-increasing function of $N_{RF}$ for $N_{RF}\le L$, then keeps unchanged for $N_{RF}\ge L$. The average training length for $N_{RF}=L$ is a lower bound for a general $N_{RF}$ value and the average training length for $N_{RF}=1$ is an upper bound.

With the above discussion, in the following analysis, we only consider the scenario of $1\le N_{RF}\le L$. Define
\begin{equation}\label{xi_ref}
\xi(i,j)\triangleq {\binom{i-1}{j}\binom{N_t-i}{L-j-1}}/{\binom{N_t}{L}}
\end{equation}
for $i=2,\cdots,N_t-1$ and $j<i$, which is the probability of a path being aligned by the $i$th beam and $j$ paths being aligned by the first $i-1$ beams. Define
\begin{eqnarray*}
&&\hspace{-3mm}\beta_{j,l} \triangleq \frac{{{{( - 1)}^l}j!{L }^{{N_{RF}}}}}{{(j - {N_{RF}} - l)!({N_{RF}} - 1)!({N_{RF}} - 2)!l!{}}}\\
&&\hspace{3cm} \text{ for } j=0,\cdots,L-1, l\le j-N_{RF},\\
&&\hspace{-3mm} b_i^{(1)} \triangleq \max\{0,L-1-N_t+i\},\ b_i^{(2)} \triangleq \min\{i-1,L-1\}  \\
&&\hspace{3cm}\text{ for } i=2,\cdots,N_t-1.
\end{eqnarray*}  The following theorem has been proved.

\begin{theorem1}\label{theorem1}
For the hybrid SU massive antenna system with $N_t$ BS antennas, the $L$-path channel, $N_{RF}\le L$ RF chains, and the target normalized received SNR $\alpha$, the average training length of the proposed IT-SU scheme is
\begin{eqnarray}\label{tr-len}
T_{\text{IT-SU}}=N_t-\sum_{i=1}^{N_t-1}(N_t-i) P_i,
\end{eqnarray}
where \begin{equation}\label{eq8}
P_1=\frac{L}{N_t}e^{\frac{- L\alpha}{N_t}},
\end{equation}
for $i=2,...,N_t-1$,
\begin{eqnarray}\label{eq_9}
&&\hspace{-1cm}P_i=
\left\{ \begin{array}{ll}
\sum\limits_{j=b_i^{(1)}}^{b_i^{(2)}} \xi(i,j)
  e^{\frac{-  L \alpha}{N_t}} \left( 1 - e^\frac{ - L \alpha }{N_t}
   \right)^j  & \text{if } N_{RF}=1,\\
\hspace{-2mm}\sum\limits_{j=b_i^{(1)}}^{\min(\hspace{-0.5mm}N_{RF}\hspace{-0.4mm}-\hspace{-0.4mm}1,b_i^{(2)}\hspace{-0.5mm})} \xi(i,j)
(\frac{L\alpha }{N_t})^j e^{\frac{ - L\alpha }{N_t}}/{j!}
 & \\
 \hspace{0.1cm}+\hspace{-2mm}\sum\limits_{j=\max(\hspace{-0.5mm}N_{RF}\hspace{-0.5mm},b_i^{(1)}\hspace{-0.5mm})}^{b_i^{(2)}}
\hspace{-0cm}
\hspace{-1mm}\xi(i,j)\hspace{-1mm}
\left( P_j^{(1)}\hspace{-1mm}+\hspace{-1mm} P_j^{(2)}\right)
  & \text{if }1<\hspace{-0.5mm}\hspace{-0.5mm}N_{RF}\hspace{-0.5mm}\le \hspace{-0.5mm}L,
\end{array} \right.
\end{eqnarray}
and ${\rm P}_j^{(1)}$ and ${\rm P}_j^{(2)}$ are shown in \eqref{P_1} and \eqref{P_2} at the top of next page.
\begin{figure*}[t!]
\begin{eqnarray}\label{P_1}
\nonumber&&{\rm P}_j^{(1)} \triangleq \left(\frac{1-N_{RF}}{L}\right)^{N_{RF}}\frac{{e^{ - \frac{{L\alpha }}{{{N_t}}}}}}{N_{RF}-1}\sum\limits_{l = 0}^{j - {N_{RF}}} {\beta_{j,l}\sum\limits_{m = 0}^{{N_{RF}} - 2} \binom{N_{RF}-2}{m} {{ {\frac{{(-1)^m}}{{(l + 1)^{{N_{RF}}}}}}}}}\\
      && \hspace{3cm}\times \frac{\left[ {{t^{m + 1}}\Upsilon \left( {{N_{RF}} - 1 - m,t} \right) + \Gamma \left( {{N_{RF}},t} \right)} \right]_{t = 0}^{t = \frac{{L\alpha  \left( {l + 1} \right)}}{{{N_t}{N_{RF}} }}}}{{m + 1}}.
\end{eqnarray}
\hrulefill
\end{figure*}
\begin{figure*}[t!]
\begin{eqnarray}\label{P_2}
\nonumber&&\hspace{-0.2cm}{\rm P}_j^{(2)} \triangleq \left(\frac{1-N_{RF}}{L}\right)^{N_{RF}}\frac{{-e^{ - \frac{{L\alpha }}{N_t}}}}{(N_{RF}-1)^2}
\sum\limits_{l = 0}^{j - {N_{RF}}} \beta_{j,l}\sum\limits_{m = 0}^{{N_{RF}} - 2} \binom{N_{RF}-2}{m}
\frac{1}{(N_{RF}-1)^m(l+1)^{N_{RF}}}\\
&&\hspace{2cm}\times \sum\limits_{n = 0}^m \binom{m}{n} \left(\frac{-L\alpha(l+1)}{N_t}\right)^n\frac{\left[ {{t^{m - n + 1}}\Upsilon \left( {{N_{RF}} - 1 - m,t} \right) + \Gamma \left( {{N_{RF}} - n,t} \right)} \right]_{\frac{L\alpha(l + 1)}{N_tN_{RF} }}^0}{{m - n + 1}}.
\end{eqnarray}
\end{figure*}
\end{theorem1}
\begin{IEEEproof}
See Appendix \ref{proofD_theorem1}.
\end{IEEEproof}

Theorem \ref{theorem1} provides an analytical expression on the average training length of the proposed IT-SU scheme. Other than the two special functions $\Upsilon$ and $\Gamma$, it is in closed-form. The two functions are well studied and their values can be easily obtained. 
The $P_i$ given in (\ref{eq8}) and (\ref{eq_9}) is the probability that the training length is $i$. Together, $P_1,\cdots,P_{N_t}$ form the probability mass function (PMF) of the training length. From (\ref{tr-len}), it can be easily concluded that the average training length of the proposed scheme is always less than $N_t$ since $P_i\ge0$, $\forall i$.

The result in Theorem \ref{theorem1} is general and applies for arbitrary values of $N_{RF},L,N_t$. 
But due to the complicated format, it is hard to see insightful behaviours of the average training length directly. In what follows, several typical scenarios for massive antenna systems are considered.


\subsubsection{The Channel with Finite $L$, i.e., $L=\mathcal{O}(1)$}
We first consider the channels for the massive antenna system with a finite number of paths. That is, $L$ is a finite value while $N_{t} \rightarrow\infty$. The asymptotic result on the average training length of the proposed IT-SU scheme is provided for both the special case of single RF chain and the general case of multiple RF chains.
\begin{lemma}\label{corollary4}
For the hybrid massive antenna system with $N_t\gg1$ BS antennas, finite constant  number of channel paths $L$, and target normalized received SNR $\alpha$, when the number of RF chains is one, i.e., $N_{RF}=1$, or is the same as the path number, i.e., $N_{RF}=L$, the average training length of the proposed IT-SU scheme can be written as follows:
\begin{equation}\label{eq16}
T_{\text{IT-SU}}=\frac{N_t}{L+1}+\mathcal{O}(1).
\end{equation}
\end{lemma}
\begin{IEEEproof}
See Appendix \ref{appendix F}.
\end{IEEEproof}

The result in Lemma \ref{corollary4} shows that for the two special $N_{RF}$ values, the average training length of the IT-SU scheme increases linearly with $N_t$, but the slope decreases linearly with $L$, the number of channel paths. The traditional NIT-SU scheme with full training has a fixed training length $N_t$. Thus the proposed IT-SU scheme has significant saving in training time as $N_t$ is very large. For example, when $N_{RF}=L=1$, we have $T_{\text{IT-SU}}\approx N_t/2$, meaning that the IT-SU scheme reduces the average training length by half. It is noteworthy that this gain in time is obtained with no cost in outage probability performance (more details will be explained in the next subsection). Moreover, the average training length is independent of the threshold $\alpha$.

From the discussion at the beginning of this subsection, we know that for any value of $N_{RF}$, the average training length is lower bounded by its value for $N_{RF}=L$ and upper bounded by its value for $N_{RF}=1$. Thus the analytical results for the two special cases in Lemma \ref{corollary4} lead to the following corollary.

\begin{corollary}
For the hybrid massive antenna system with $N_t\gg1$ BS antennas, finite constant number of channel paths $L$, and target normalized received SNR $\alpha$, the average training length of the proposed IT-SU scheme can be written as \eqref{eq16} for any number of RF chains $N_{RF}$.
\end{corollary}

\subsubsection{The Channel with Linearly Increasing $L$, i.e., $L=\mathcal{O}(N_t)$}\label{section_cNt}
The next typical scenario to consider is that the number of channel paths has a linear scaling with the number of BS antennas. That is, $L=cN_t$ while $N_{t} \rightarrow\infty$ for a constant but arbitrary $c$. For this case, due to the more complicated form of $P_i$ than that of the finite $L$ case, simple expression of the average training length is hard to obtain. Two special cases are analyzed in what follows.

For the special case where $N_{RF}=1$ and $c=1$, i.e., single RF chain and i.i.d.~channels, we have $b_i^{(1)}=b_i^{(2)}=i-1$ for $i=2,...,N_t-1$. Thus, from \eqref{eq8} and \eqref{eq_9},
\[P_i=e^{-\alpha}(1-e^{-\alpha})^{i-1}\] for $i=1,...,N_t-1$. This is the same as the result of the interleaved antenna selection scheme for full-digital massive antenna systems with i.i.d.~channels in \cite{Koyuncu_interleaving}.
The same asymptotic upper bound on the average training length for large $N_t$ can be obtained as:
\[
T_{\text{IT-SU}} = e^{\alpha}(1-(1-e^{-\alpha})^{N_t})\rightarrow e^{\alpha}\ \text{when } N_t\rightarrow \infty.
 \]
This bound is only dependent on the threshold $\alpha$.

For another special case where $N_{RF}=L$ and $c=1$, from \eqref{eq8} and \eqref{eq_9}, we have $P_i=e^{-\alpha}{\alpha}^{i-1}/(i-1)!$ for $i=1,...,N_t-1$. This is in accordance with that of the interleaved Scheme D for full-digital massive antenna systems with i.i.d.~channels in \cite[Theorem 2]{Koyuncu_interleaving}. The corresponding upper bound on the average training length is $1+\alpha$, which again is only dependent on the threshold $\alpha$.

\begin{figure}[htb]
\centering
\includegraphics[scale=0.49]{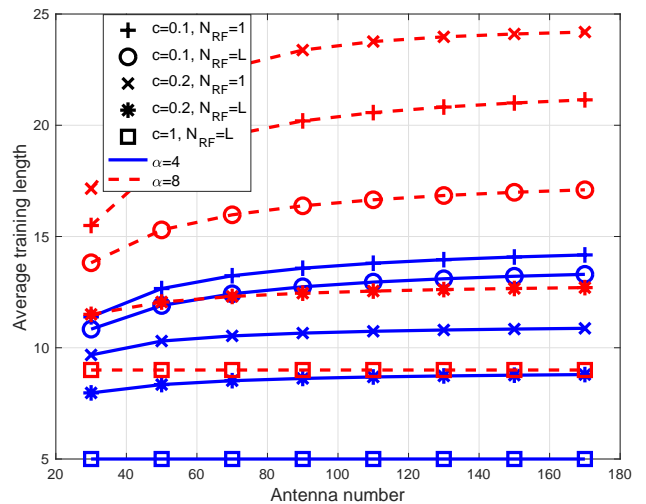}
\captionsetup{margin=5pt,font=small}
\caption{Average training length of the IT-SU scheme for $\alpha=4, 8$.}\label{L_Ntc_study2}
\end{figure}

For the general case of $0<c<1$, the expression of $P_i$ in \eqref{eq_9} can be used along with numerical evaluations for further studies. Fig.~\ref{L_Ntc_study2} shows the average training length $T_{\text{IT-SU}}$ with respect to $N_t$ for different parameter values.
The following observations are obtained from the plot.

\begin{itemize}
  \item For any $N_{RF}$ value, $T_{\text{IT-SU}}$ asymptotically approaches a constant upper bound that is independent of $N_t$.
  \item $T_{\text{IT-SU}}$ is an increasing function of the threshold $\alpha$. Thus the advantage of the proposed scheme over NIT-SU degrades for larger $\alpha$. The advantage degradation with increasing $\alpha$  is most severe when $c=1$ and $N_{RF}=1$, but for larger $N_{RF}$ or smaller $c$, it is considerably slower.
  \item When $N_{RF}=L$, $T_{\text{IT-SU}}$ is a decreasing function of $c$. When $N_{RF}=1$, depending on the value of $\alpha$, $T_{\text{IT-SU}}$ may not be a monotonic function of $c$. This can be explained by the two opposite effects of increasing $c$: the increase in multi-path diversity and the decrease in average path power. For $N_{RF}=L$, there are enough beams to be used to compensate for the second effect, thus larger $c$ tends to decrease $T_{\text{IT-SU}}$ via higher multi-path diversity. For $N_{RF}=1$, the second effect is more dominant, thus larger $c$ tends to increase $T_{\text{IT-SU}}$ due to the path power loss, especially for large threshold $\alpha$.
\end{itemize}

\subsection{Outage Performance Analysis}
\label{outage_int}
In this subsection, the outage probability of the proposed IT-SU scheme is analyzed.
\begin{theorem1}\label{outage_theorem}
For the hybrid SU massive  antenna system with $N_t$ BS antennas, the $L$-path channel, $N_{RF}\le L$ RF chains and the target normalized received SNR $\alpha$, the outage probability of the IT-SU scheme is
\begin{eqnarray}\label{out_ITsu1}\nonumber
\hspace{-0.7cm}&&{\rm out}(\text{IT-SU})=\\
\nonumber\hspace{-0.7cm}&&\binom{L}{N_{RF}}\left[{\frac{{\Upsilon \left( {{N_{RF}},\frac{{\alpha L}}{{{N_t}}}} \right)}}{{({N_{RF}} - 1)!}}}
+\hspace{-2mm}\sum\limits_{l = 1}^{L - {N_{RF}}} \hspace{-1mm}{{{( - 1)}^{{N_{RF}} + l - 1}}}\binom{L\hspace{-0.5mm}-\hspace{-0.5mm}N_{RF}}{l}\right.\\
\hspace{-0.7cm}&&
\left.\times{{\left( {\frac{{{N_{RF}}}}{l}} \right)}^{{N_{RF}} - 1}}
{\left( {\frac{{{e^{\left( { - 1 - \frac{l}{{{N_{RF}}}}} \right)\frac{{\alpha L}}{{{N_t}}}}} - 1}}{{\left( { - 1 - \frac{l}{{{N_{RF}}}}} \right)}} - B(l)} \right)}\right],
\end{eqnarray}
where
\begin{equation}\label{out_ITsu2}
B(l) \triangleq \left\{ {\begin{array}{ll}
\hspace{-2mm}{\sum\limits_{m = 0}^{{N_{RF}} - 2} {\frac{1}{{m!}}{{\left( { - \frac{l}{{{N_{RF}}}}} \right)}^m}} \Upsilon \left( {m + 1,\frac{{\alpha L}}{{{N_t} }}} \right)} & N_{RF}\ge 2\\
0 & \text{otherwise}
\end{array}.} \right.
\end{equation}
\label{thm-SU-out}
\end{theorem1}
\begin{IEEEproof}
See Appendix \ref{appA_Lemma1}.
\end{IEEEproof}

Theorem \ref{thm-SU-out} provides an analytical expression for the outage probability of the proposed IT-SU scheme. The expression is in closed-form other than the special function $\Upsilon$. 
Although the effect of $N_{RF}$ on the outage performance of the IT-SU scheme is implicit in \eqref{out_ITsu1}, from its derivations, we have  ${\rm out}({\text{IT-SU}})={\rm Pr}(x\le\alpha/N_t)$ where $x$ is the sum of the largest $N_{RF}$ elements in $\{\|{\bar h}_{i}\|^2,i\in {\mathcal I}\}$. Apparently, for any channel realization, the value of $x$
increases as $N_{RF}$ increases from $1$ to $L$. Therefore, for any given finite $\alpha$, larger $N_{RF}$ means smaller outage probability.

\subsubsection{Single RF Chain Analysis}
To obtain further insights in the effect of $N_t$, $L$ and $\alpha$ on the outage performance, we consider  the special case with $N_{RF}=1$ in what follows.
\begin{lemma}\label{Corollary 1}
For the hybrid massive antenna system with $N_t$ BS antennas, the $L$-path channel, single RF chain, and the target normalized received SNR $\alpha$, the outage probability of the IT-SU scheme is as follows:
\begin{equation}
{\rm out}({\text{IT-SU}})=(1-e^{\frac{-\alpha L}{N_t}})^{L}.
\label{outP-1}
\end{equation}
\end{lemma}
\begin{IEEEproof}
See Appendix \ref{appendB_Corollary1}.
\end{IEEEproof}

It can be seen from (\ref{outP-1}) that for arbitrary values of $\alpha$, $P$ and arbitrary scaling of $L$ with respect to $N_t$ between constant and linear, i.e., $L=\mathcal{O}(N_t^{r_L})$ for $r_L\in[0,1]$, we have
\[\lim_{N_t\rightarrow \infty}{\rm out}({\text{IT-SU}})=0.
\]
This means that for the SU massive antenna system with single RF chain, arbitrarily small outage probability can be obtained  for any desired date rate and any fixed power consumption $P$ as long as $N_{t}$ is large enough. This shows the advantage of having massive antenna array at the BS.

Specifically, for finite $L$, i.e., $L=\mathcal{O}(1)$, we have
\[
{\rm out}(\text{IT-SU})= \left(\frac{\alpha L}{N_t}\right)^L +\mathcal{O}\left(N_t^{-(L+1)}\right),
\]
meaning that the outage probability scales as $\mathcal{O}\left(N_t^{-L}\right)$ for large $N_t$. For linearly increasing $L$ where $L=cN_t$, we have
\[
{\rm out}(\text{IT-SU})= (1-e^{-\alpha c})^{cN_t},
\]
meaning that the outage probability decreases exponentially with respect to $N_t$. For i.i.d.~channels where $c=1$, the outage probability of the IT-SU scheme reduces to $(1-e^{-\alpha})^{N_t}$. This is the same as that of the antenna selection scheme in the full-digital massive antenna systems with i.i.d.~channels \cite{Koyuncu_interleaving}.

\subsubsection{Multiple RF Chain Analysis}\label{sec_MRF}
For the case of multiple RF chain where $1<N_{RF}\le L$, since larger $N_{RF}$ results in smaller outage probability, for both finite $L$ and $L=cN_t$, it can be concluded that arbitrarily small outage probability can also be achieved for an arbitrary date rate with any fixed power consumption $P$ when $N_t$ is large enough.

\subsubsection{Comparison with NIT Schemes}\label{section_partial_com}
In Section \ref{sec-SU-time}, the proposed IT-SU scheme was compared with the NIT-SU scheme with full training in terms of training length. Here, we give the outage probability comparison. As utilized in the proof of Theorem \ref{thm-SU-out}, for the IT-SU scheme, an outage happens only when all beams have been trained and no beam combination can satisfy the target SNR requirement. This is the same as that of the NIT-SU scheme with full training, thus the two schemes have the same outage performance.

Another possible non-interleaved scheme is to have partial training with a fixed training length of $L_t<N_t$ \cite{Lee_Impact}. It can be seen easily that as $L_t$ decreases, the outage probability of the  partial non-interleaved scheme increases. Thus, the proposed IT-SU scheme is superior in terms of outage probability compared with the NIT-SU scheme with the same training length. Numerical validation will be given in the simulation section.

\section{Interleaved Training for MU Transmission}
This section is on the more general and complicated MU systems, where the joint beam-based interleaved training and the corresponding MU transmission is proposed. Compared to SU systems where the optimal transmission is the maximum-ratio combining of the best trained beams, the beam assignment problem is a challenging but crucial part of the MU transmission. In what follows, we first study the beam assignment issue. Subsequently, the joint beam-based interleaved training and MU transmission is proposed.

\subsection{Feasible Beam Assignment and MU Beam Assignment Methods}
\label{sec-MU-BA}
For the hybrid massive antenna BS to serve multiple users with a fixed beam codebook, a typical idea is to assign a beam to each user. But the beam assignment problem is far from trivial and is a dominant factor of the performance. We first introduce the definition of feasible beam assignment, then propose MU beam assignment methods.

\begin{defn}\label{defn1}
For the hybrid massive antenna downlink  serving $U$ users with codebook-based beam transmission, a \textit{beam assignment} is an ordered $U$-tuple, $(n_1,\dots,\dots n_U)$, where $n_i$ is the index of the beam assigned for User $i$. A \textit{feasible beam assignment} is a beam assignment where the resulting effective channel matrix as given in (\ref{effective channel}) has full rank.
\end{defn}

In other words, a beam assignment is feasible if ZF baseband can be conducted with no singularity, thus the received SNR, denoted as ${\rm SNR}_{\rm MU}$ for the MU transmission, or $\lambda$ in \eqref{mu_SINR} is non-zero.
If an infeasible beam assignment is used for the MU transmission, ZF baseband precoding cannot be conducted. Even if other baseband precoding, e.g., regularized ZF, is used, the received SINR will be very small due to the high interference
and outage occurs.
Thus feasible beam assignment is a necessary condition to avoid outage. Two cases that can cause infeasible beam assignment are 1) all beams $(n_1,\dots,\dots n_U)$  are zero-beams for any user thus the effective channel matrix has a row with all zeros, and 2) one beam is assigned to more than one user thus two identical columns appear in the effective channel matrix. On the other hand, depending on the effective channel values and interference level, a feasible beam assignment may or may not be able to avoid outage.

A straightforward and optimal beam assignment method is the exhaustive search. Denote the set of known (e.g., already trained) non-zero beam indices for User $u$ as $\mathcal{B}_u$. Define $\mathcal{B} \triangleq \cup_{u=1}^U \mathcal{B}_u$. By searching over all possible feasible beam assignments over $\mathcal{B}$ and finding the one with the maximum $\lambda$, the optimal beam assignment is obtained. The complexity of the exhaustive search is however $\mathcal{O}(|\mathcal{B}|^U)$, which is unaffordable for large $|\mathcal{B}|$ and/or $U$.

Thus for practice implementation, beam assignment methods with affordable complexity are needed. To serve this purpose, we transform the SNR maximization problem for the beam assignment to the problem of maximizing the minimum effective channel gain among the users, i.e.,  \begin{eqnarray}
\arg\max_{n_1,\cdots,n_U\in \mathcal{B}} \min_u \{|{\bar h}_{u, n_u}| \}.
\label{max-min}
\end{eqnarray}
Then by drawing lessons from the extended optimal relay selection (ORS) method for MU relay networks in \cite{Atapattu_Relay}, the following beam assignment algorithm is proposed. First, the original ORS algorithm in \cite{Sharma_Optimal} is used to maximize the minimum effective channel gain. Suppose that the minimum gain is with User $i$ and Beam $j$. Then we delete User $i$ and Beam $j$ and apply the original ORS scheme again to the remaining users and beams. This procedure is repeated until all users find their beams. It has been shown in \cite{Atapattu_Relay,Sharma_Optimal} that this scheme not only achieves an optimal solution for (\ref{max-min}), but also achieves the unique optimal solution that maximizes the $u$th minimum channel gain conditioned on the previous $1$st to the $(u-1)$th minimum channel gains for all $u$. Further, the worst case complexity of this scheme is $\mathcal{O}(U^2|\mathcal{B}|^2)$, much less than that of the exhaustive search. This beam assignment is referred to as the \textit{max-min assignment}. With respect to the outage probability performance, it is suboptimal. But the method targets at maximizing the diagonal elements of the effective channel matrix, which in general is beneficial to ZF transmission. Simulation results exhibited in Section \ref{sec_simu} show that its outage performance loss is small compared with the exhaustive search especially for channels with small $L$.

\subsection{Joint Beam-Based Interleaved Training and MU Transmission Design}
Similar to the SU case, the main goal of the interleaved training and joint MU data transmission scheme (referred to as the IT-MU scheme) is to save training time while preserving the outage probability
performance. The fundamental idea is to conduct the training of each beam sequentially and terminate right after enough beams have been trained to avoid outage. However, different from the SU case, a big challenge for the MU case is that each user does not know the effective channels of other users since user cooperation is not considered. Thus the users are not able to decide whether to terminate the training interval given an SNR threshold. Our solution for this is to make users feed back their acquired non-zero effective channels during training and let the BS to make the decision. Other differences of the MU scheme to the SU one include the initial training steps, the beam assignment problem, and the termination condition for the training interval. These will be studied in details in the explanation of the scheme that follows.

\begin{algorithm}[htb]
\caption{The Joint Beam-Based Interleaved Training and MU Transmission (IT-MU) Scheme.}
\label{IMTS}
\begin{algorithmic}[1]
\STATE The BS trains the $1$st to $U$th beams. User $u, \forall u$ estimates the corresponding effective channels $\bar{h}_{u,1},...,\bar{h}_{u,U}$ and constructs its set of non-zero beam indices $\mathcal{B}_u$;
\STATE If $|\mathcal{B}_u|=0$, User $u, \forall u$ feeds back ``0". Otherwise, User $u$ feeds back the non-zero effective channel values along with their beam indices;
\IF {any user's feedback is ``0" or $|\mathcal{B}|<U$}
\STATE set $os=1$ and \textit{goto Step 13};
\ELSE
\STATE The BS finds a beam assignment on $\mathcal{B}$;
\IF {the beam assignment is not feasible or the resulting received SNR is below the outage threshold}
\STATE set $os=1$ and \textit{goto Step 13};
\ELSE
\STATE set $os=0$ and \textit{goto Step 27};
\ENDIF
\ENDIF
\FOR{$i=U+1,...,N_t$}
\STATE The BS trains the $i$th beam; User $u, \forall u$ estimates the corresponding effective channel $\bar{h}_{u,i}$;
\STATE For all $u$, if $\|\bar{h}_{u,i}\|=0$, User $u$  feeds back ``0"; else  User $u$ feeds back the value $\bar{h}_{u,i}$ and let $\mathcal{B}_u=\mathcal{B}_u\cup \{i\}$ and $\mathcal{B}=\mathcal{B}\cup \{i\}$;
\IF { all users' feedbacks are ``0"  or $|\mathcal{B}_u|=0$ for any $u$ or $|\mathcal{B}|<U$}
\STATE set $os=1$ and \textit{continue};
\ELSE
\STATE The BS finds a beam assignment on $\mathcal{B}$;
\IF {the beam assignment is not feasible or the resulting received SNR is below the  outage threshold}
\STATE set $os=1$ and \textit{continue};
\ELSE
\STATE set $os=0$ and \textit{goto Step 27};
\ENDIF
\ENDIF
\ENDFOR
\IF {$os=0$}
\STATE The BS uses the found beam assignment to construct ${\bf F}_{RF}$ and the ZF ${\bf F}_{BB}$ for MU transmission;
\ENDIF
\end{algorithmic}
\end{algorithm}

The detailed steps for the proposed IT-MU scheme is given in Algorithm \ref{IMTS}. At the beginning of this scheme, the first $U$ beams in the codebook are trained and every user estimates the corresponding effective channels and constructs its set of non-zero beam indices $\mathcal{B}_u$. Then the non-zero beam values and their indices are fed back to the BS (with this information, the BS also knows $\mathcal{B}_u, \forall u$). While for the SU case the beams are trained one by one, $U$ beams need to be trained initially for the MU case since at least $U$ beams are needed for a feasible beam assignment.

After this initial training stage, if for any user, no non-zero beam is found (in which case the user feeds back ``0'') or $|\mathcal{B}|<U$ where $\mathcal{B}$ is the union of non-zero beam indices of all users, the training of the next beam starts. Otherwise, the BS finds a beam assignment on $\mathcal{B}$ with either the exhaustive search or the max-min method given in Section \ref{sec-MU-BA}. If the beam assignment is feasible and can avoid outage, training terminates and data transmission starts with this beam assignment and the corresponding ZF baseband precoding as shown in Section \ref{HP-OP}.
Otherwise, the BS starts the training of the next beam. When the new $i$th beam has been trained, each user again estimates the corresponding effective channel. If it is a zero-beam, the user feeds back ``0''; otherwise, it feeds back the effective channel value.
If this $i$th beam is a zero-beam for all users or any user still has no non-zero beam or the updated $|\mathcal{B}|$ is still less than $U$, the BS starts the training of the next beam if an un-trained beam is available. Otherwise, the BS finds a beam assignment on $\mathcal{B}$ with either the exhaustive search or the max-min method. If the beam assignment is feasible and can avoid outage, training terminates and transmission starts.
Otherwise, the BS starts the training of the next beam if an un-trained beam is available. The procedure continues until a beam assignment that can avoid outage is found or there is no new beam for training.

\subsection{Discussion on Average Training Length and Outage Performance}
For the IT-MU scheme, the minimum possible training length is $U$ and the maximum possible training length is $N_t$. Similar to the IT-SU scheme, it is reasonably expected that the IT-MU scheme has a smaller average training length than the NIT-MU scheme with full training. Meanwhile, since complete effective CSI is available for the BS if necessary in the IT-MU scheme, it achieves the same outage performance. Moreover, the outage probability of the IT-MU scheme is smaller than that of the NIT-MU scheme with partial training at the same training length. 

\section{Numerical Results and Discussions}\label{sec_simu}
In this section, simulation results are shown to verify the analytical results in this paper. Meanwhile, properties of the proposed interleaved training and joint transmission schemes are demonstrated. We also make comparison with non-interleaved schemes. 

\begin{figure*}[htb]
  \normalsize
  \centering
  \hspace{-0.5cm}
  \begin{minipage}[t]{0.49\textwidth}
    \centering
    \includegraphics[scale=0.50]{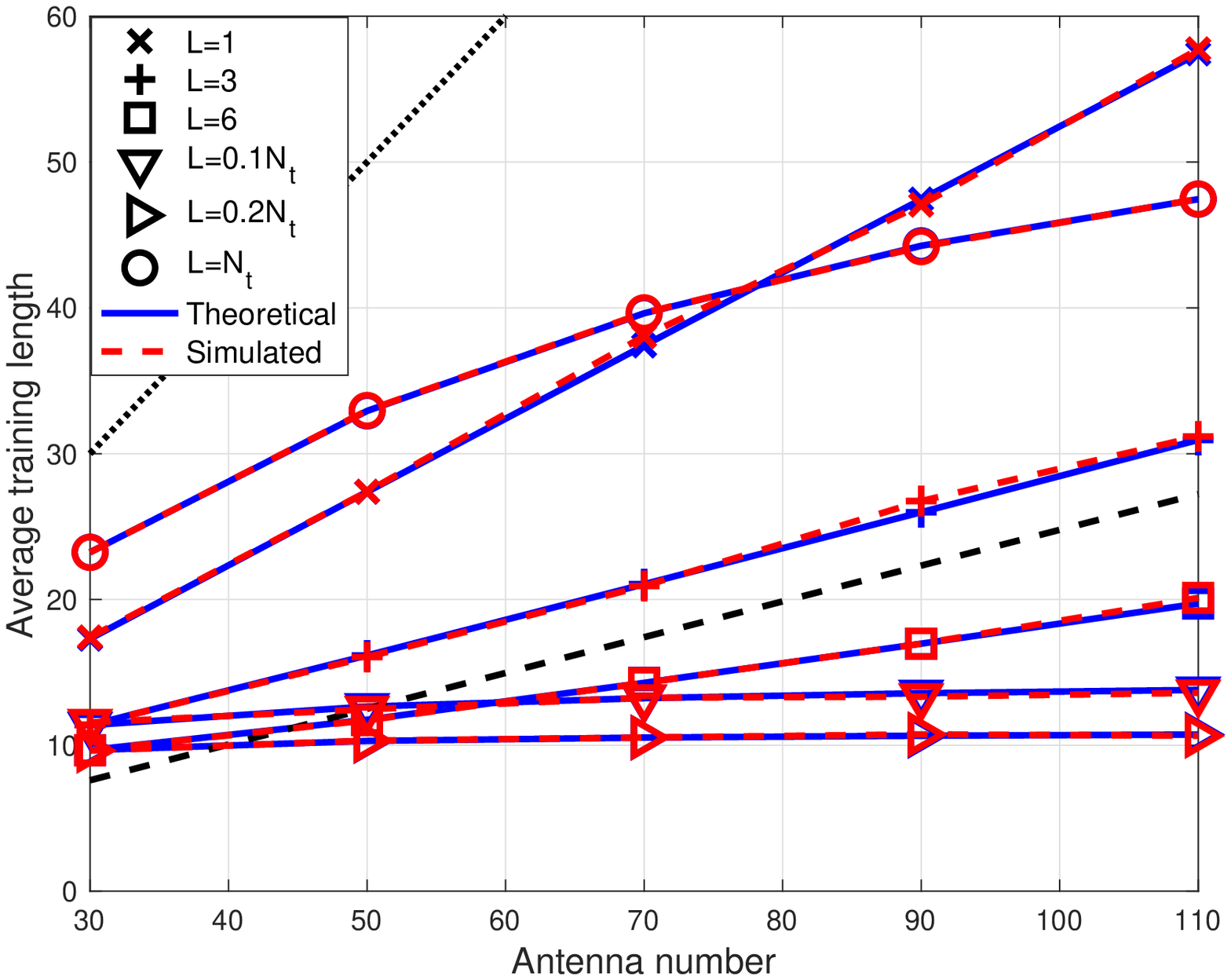}\vspace{-1mm}
  \end{minipage}
  \hspace{0.2cm}
  \begin{minipage}[t]{0.49\textwidth}
    \centering
    \includegraphics[scale=0.5]{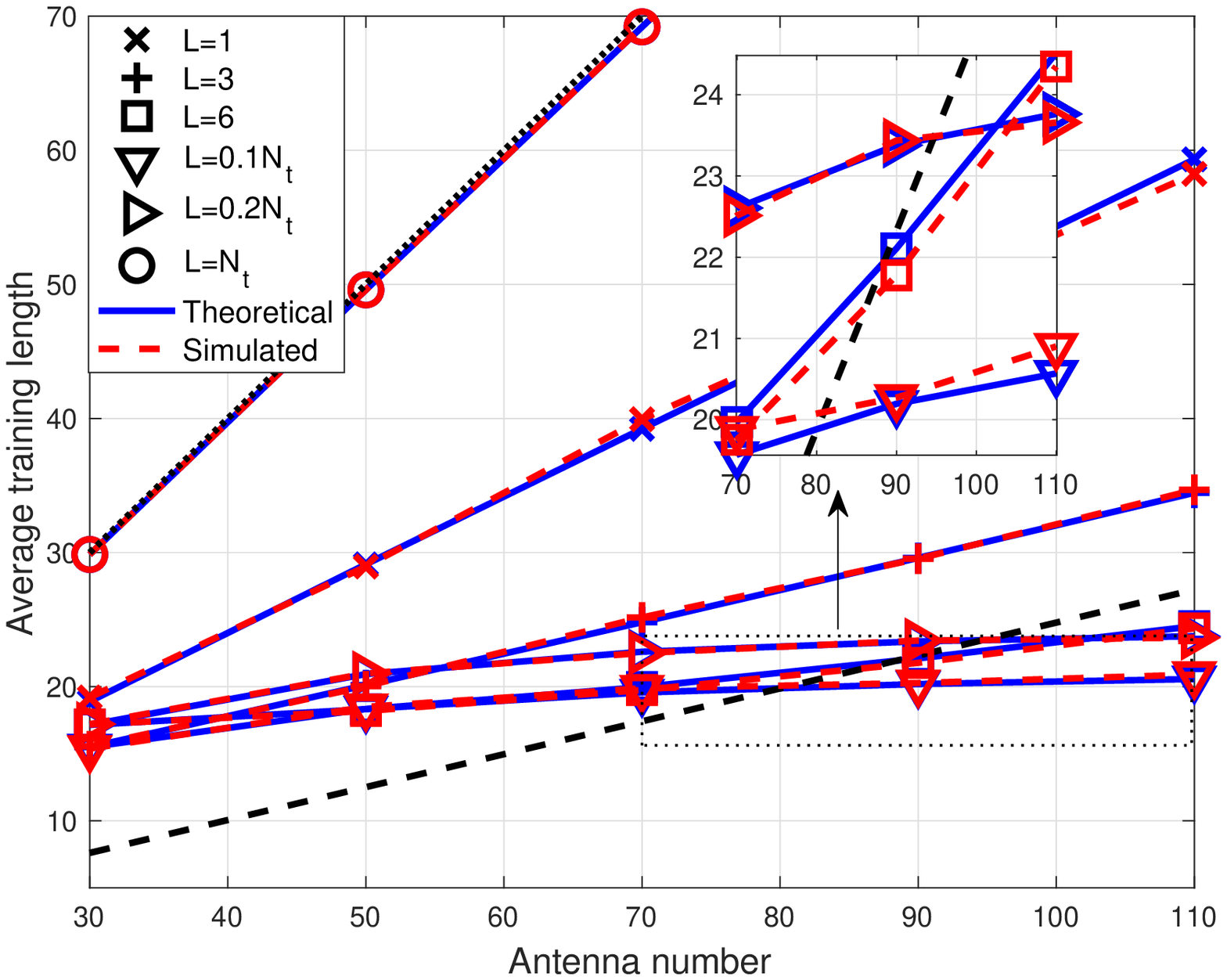}\vspace{-1mm}
  \end{minipage}
      \caption{Average training length of the IT-SU scheme with $N_{RF}=1$ for $\alpha=4$ (left) and $\alpha=8$ (right). The dotted line shows the training length of the NIT-SU scheme with full training.}
    \label{fig_training_rate_Nrf1}
\end{figure*}

In Fig.~\ref{fig_training_rate_Nrf1}, the average training length of the IT-SU scheme in Algorithm \ref{ISTS} is shown for $N_{RF}=1$ and $\alpha=4, 8$\footnote{For $P=0$ dB, $\alpha=4, 8$ mean that $R_{th} = 2.3219, 3.1699$ bps/Hz respectively which represent the low rate scenario.  For $P=10$ dB, $\alpha=4, 8$ mean that $R_{th} = 5.3576,  6.3399$ bps/Hz respectively which represent
the intermediate to high rate scenario.}. First, it can be seen that
 the derived average training length in Theorem \ref{theorem1} well matches the simulation. Second, for $L=1,3,6$ and $\alpha=4$, the average training length increases linearly with $N_t$ with slope about $0.50$, $0.26$ and $0.13$, respectively. These match the theoretical results in Lemma \ref{corollary4} where the slope is  $1/(L+1)=0.5,0.25,0.14$, respectively. Further, the dashed line without marker is the line of  $T_{\text{IT-SU}}={N_t}/{4}$, which is the asymptotic average training length for $L=3$ in \eqref{eq16}.
Third, for $L=c{N_t}$ where $c=0.1, 0.2$, the average training lengths approach to constants as $N_t$ increases. While for $c=1$, the asymptotic constant upper bound is less explicit since $N_t$ is not large enough to reveal the asymptotic bound $e^{\alpha}$. When $c=1$ and $\alpha=8$, the average training length is almost the same as the NIT-SU scheme with full training (dotted line) due to the high SNR requirement and limited simulation range of $N_t$. Finally, the average training length increases with $\alpha$ (much less significant for finite $L$ and small $c$) which is in accordance with the comments on Lemma \ref{corollary4} and Fig. \ref{L_Ntc_study2}.

\begin{figure*}[htb]
  \normalsize
  \centering
    \hspace{-0.5cm}
  \begin{minipage}[t]{0.49\textwidth}
    \centering
    \includegraphics[scale=0.5]{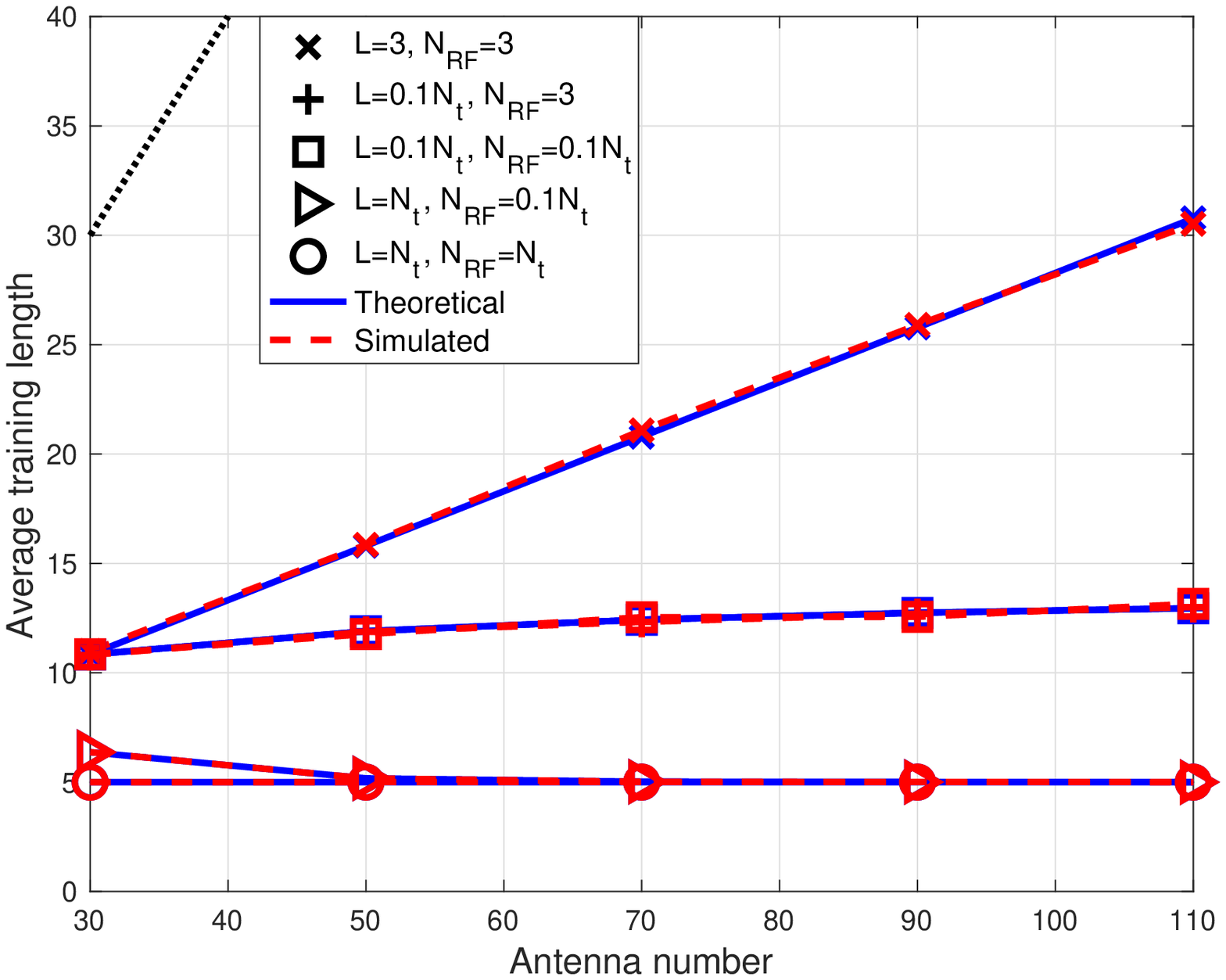}\vspace{-1mm}
    \label{training_length_Nfg}
  \end{minipage}
    \hspace{0.2cm}
  \begin{minipage}[t]{0.49\textwidth}
    \centering
    \includegraphics[scale=0.5]{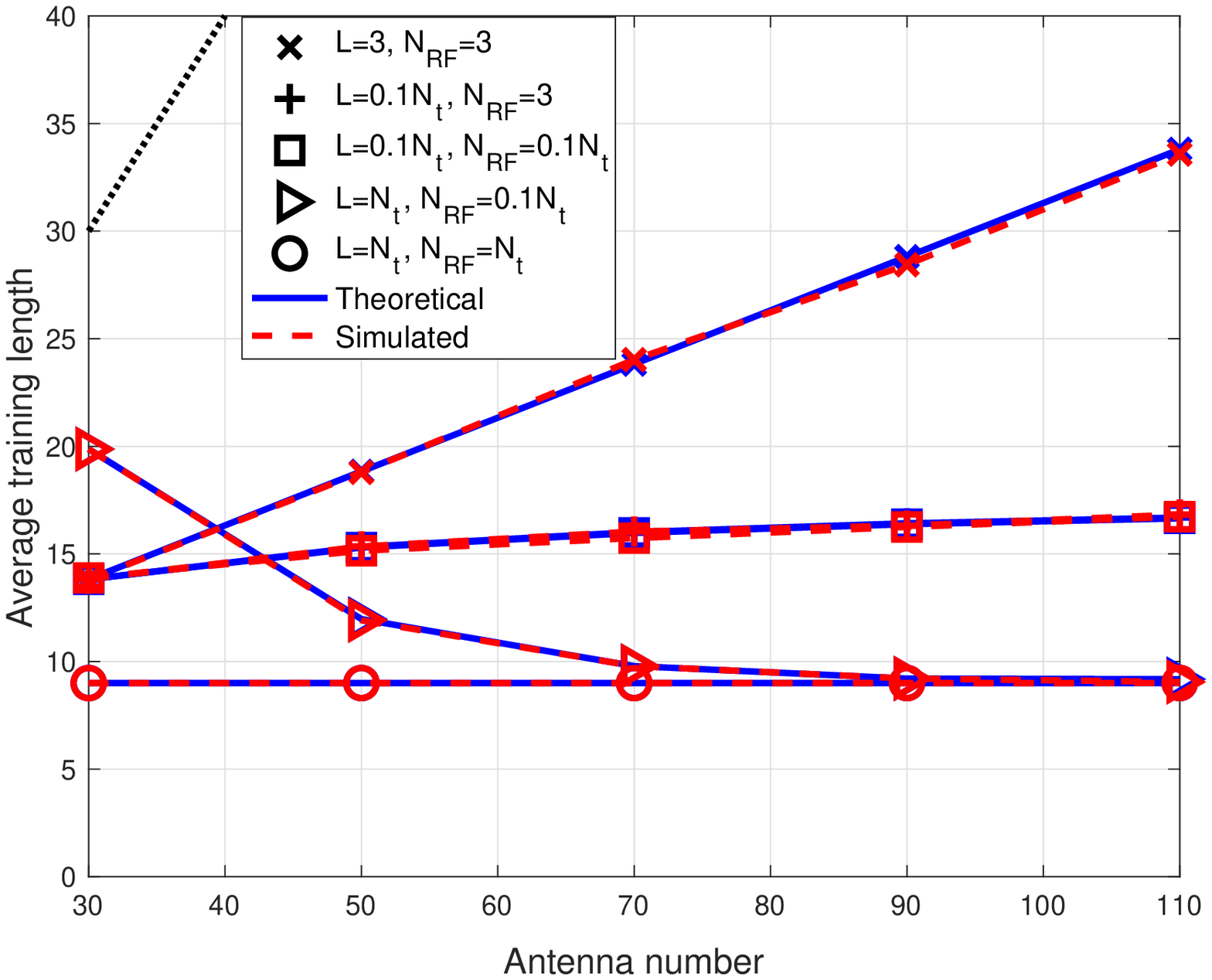}\vspace{-1mm}
    \label{training_length_Nfg_1}
  \end{minipage}
   \caption{Average training length of the IT-SU scheme with $N_{RF}>1$ for $\alpha=4$ (left) and $\alpha=8$ (right). The dotted line shows the training length of the NIT-SU scheme with full training.}\label{training_rate_gen}
   \vspace{-0.4cm}
\end{figure*}

In Fig.~\ref{training_rate_gen}, the average training length of the IT-SU scheme is studied for $N_{RF}>1$ and $\alpha=4, 8$. Again, the results in Theorem \ref{theorem1} have tight match with the simulation.
For $L=3$ and $N_{RF}=3$, the average training length increases linearly with $N_t$ with ratio about $1/(L+1)=0.25$ for $\alpha=4, 8$. Meanwhile, the average training length has negligible reduction compared with that of $N_{RF}=1$ in Fig. \ref{fig_training_rate_Nrf1}, e.g., $34.5$ for $N_{RF}=1$ and $33.8$ for $N_{RF}=3$ with $N_t=110$ and $\alpha=8$. These validate Lemma \ref{corollary4}. Further for $L=cN_t$ where $c=0.1, 1$, the average training lengths are upper bounded by different constants as $N_t$ grows. The effect of $c$, $\alpha$ and $N_{RF}$ on the upper bound has been studied in Section \ref{section_cNt}, which can be referred to directly.
Finally, the result for $L=N_{RF}=N_t$, i.e., full-digital massive MIMO with i.i.d. channels matches the theoretical result $1+\alpha$. Both Figs.~\ref{fig_training_rate_Nrf1} and \ref{training_rate_gen} show that compared with the NIT-SU scheme with full training (represented by dotted lines), the IT-SU scheme achieves huge reduction in training length with the same outage performance.

\begin{figure}[htb]
\hspace{-0.5cm}
\includegraphics[scale=0.5]{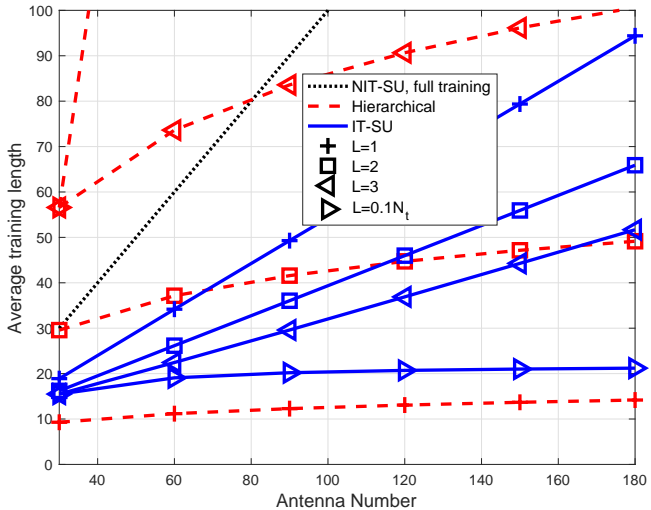}
\captionsetup{margin=5pt,font=small}
\caption{Average training length comparison of the IT-SU scheme, the NIT-SU scheme with full training, and the hierarchical scheme. $N_{RF}=1$, $\alpha=8$, and $M=3$.}\label{Response_Fig_comp_hier}
\end{figure}

In Fig.~\ref{Response_Fig_comp_hier}, comparison on the average training length is shown for the proposed scheme, the non-interleaved scheme with full training, and the hierarchical scheme with $M=3$ where $N_{RF}=1$ and $\alpha=8$. Note that for the considered setting, $M=3$ results in the minimum training length thus is the most favourable for the hierarchical scheme.
The figure shows that 1) when $L=1$, the hierarchical scheme has the lowest average training length, 2) for $L= 2$, the IT-SU scheme has lower training length than the hierarchical scheme when $N_t< 120$ and larger training length when $N_t\ge 120$, 3) for $L=3$, the IT-SU scheme has lower training length than the hierarchical scheme for all simulated values of $N_t$ in the range $[30,180]$, 4) for $L=0.1N_t$, the training length of the proposed IT-SU scheme is lower  and bounded by a constant while the hierarchical scheme experiences fast increase in the training length with larger $N_t$.

\begin{figure}[hbt]
\hspace{-0.5cm}
\includegraphics[scale=0.50]{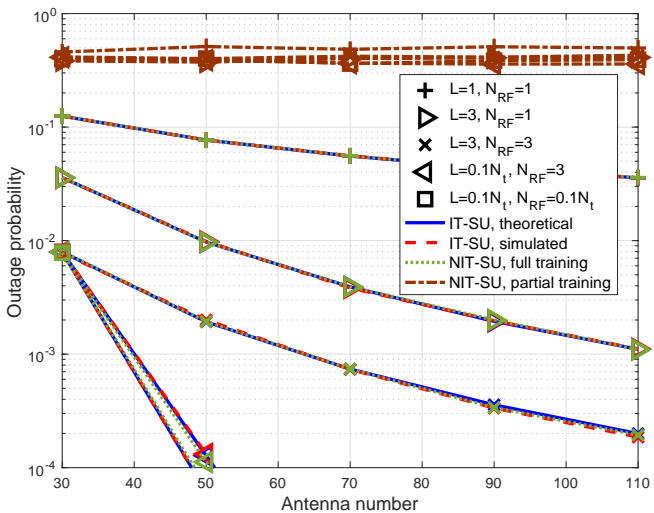}
\captionsetup{margin=5pt,font=small}
\caption{Comparison of outage probability between the IT-SU scheme and the NIT-SU scheme.}\label{op_comp}
\end{figure}

In Fig.~\ref{op_comp}, 
the outage performance of the IT-SU scheme is compared with that of the NIT-SU scheme with full training and partial training at the same training length (by setting $L_t$ to be the same as the average training length of the IT-SU scheme). The cases of $L=1,3,0.1N_t$, $N_{RF}=1,3,0.1N_t$ and $\alpha=4$ are studied. It can be seen that
1) the theoretical outage probabilities of the IT-SU scheme in Theorem \ref{outage_theorem} match the simulated values well;
2) the outage probability of the IT-SU scheme is the same as that of the NIT-SU scheme with full training, and significantly lower than that of the NIT-SU scheme with partial training;
3) the outage probability of the IT-SU scheme diminishes fast as $N_t$ grows; 4) by increasing $N_{RF}$ from $1$ to $3$ for $L=3$ or from $3$ to $0.1N_t$ for $L=0.1N_t$, the outage probability of the IT-SU scheme decreases.
These validate the discussions in Section \ref{outage_int}.

\begin{figure*}[htb]
  \normalsize
  \centering
  \hspace{-0.5cm}
  \begin{minipage}[t]{0.49\textwidth}
    \centering
    \includegraphics[scale=0.5]{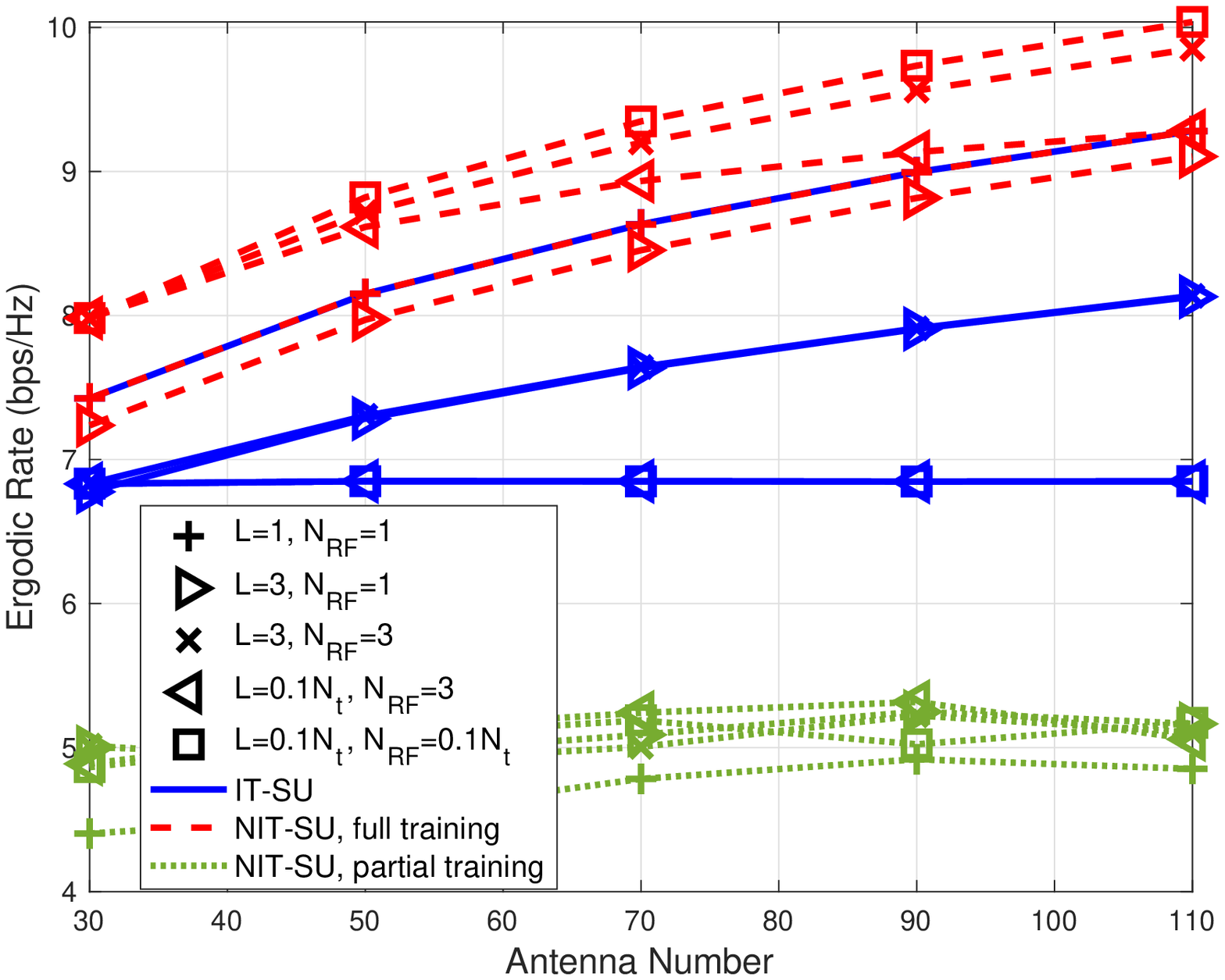}\vspace{-1mm}
  \end{minipage}
  \hspace{0.2cm}
  \begin{minipage}[t]{0.49\textwidth}
    \centering
    \includegraphics[scale=0.5]{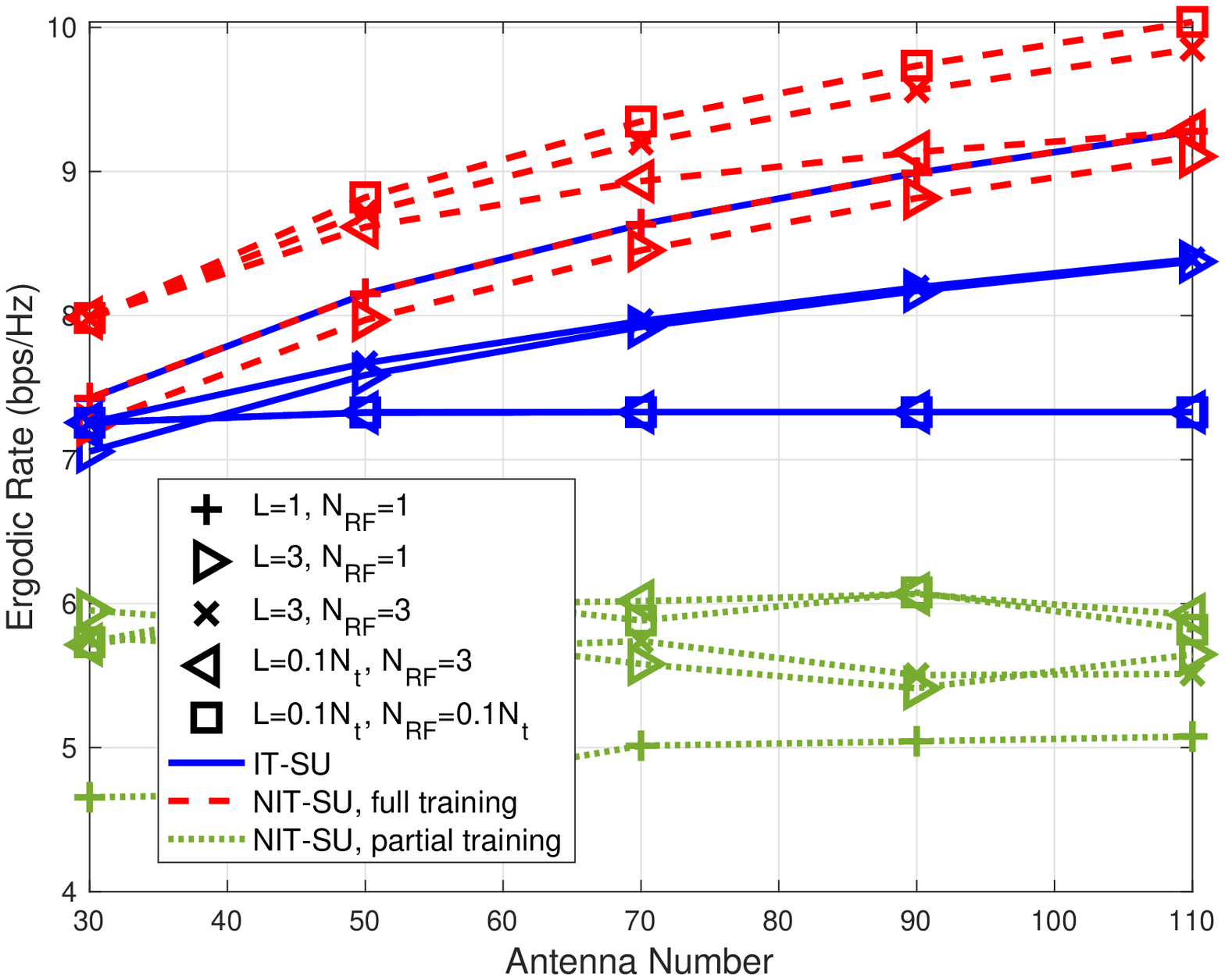}\vspace{-1mm}
  \end{minipage}
\vspace{-0.1cm}
      \caption{Ergodic rate comparison of the IT-SU scheme and the NIT-SU scheme with full/partial training for $\alpha=4$ (left) and $\alpha=8$ (right) with $P=10$ dB.}
    \label{ergodic rate comp}
\vspace{-0.7cm}
\end{figure*}

In Fig. \ref{ergodic rate comp} we show the ergodic rate of the proposed IT-SU scheme with slight modification\footnote{For the rate comparison, we change the condition in Line 5 of Algorithm 1 to ``$\|{\bar h}_{i}\|=0$ or $\sum_{l\in \mathcal{S}}\|{\bar h}_{l}\|^2 \le \alpha/N_t$ and $i<N_t$''. The only difference is for the case that all beams have been trained and an outage is still unavoidable. Previously, no transmission is conducted since outage is not avoidable, while with the change, the user feeds back the channel coefficients of the $\min(N_{RF},L)$ known non-zero beams with the largest norms and the BS uses this information for hybrid beamforming. This change has no effect on the outage performance of the proposed scheme but is sensible when considering the rate performance.} and the NIT-SU scheme for $\alpha=4, 8$ and $P=10$ dB. 
It can be seen that  the IT-SU scheme has lower ergodic rate compared with the NIT-SU scheme with full training. Compared with the NIT-SU scheme with the same average training length (partial training), the IT-SU scheme achieves higher ergodic rate. This is a very positive result for the proposed IT-SU scheme designed with the outage performance goal. It shows that interleaved and adaptive training design in general benefits the system performance compared to non-interleaved training. 

\begin{figure}[htb]
\hspace{-0.3cm}
\includegraphics[scale=0.50]{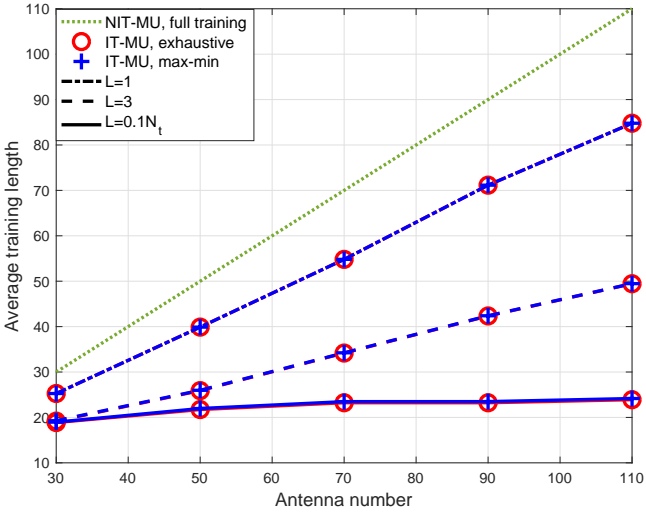}
\captionsetup{margin=5pt,font=small}
\caption{Average training length of the IT-MU scheme for $N_{RF}=U=3$ and $\bar \alpha=6$.}\label{mu_tr_U3_alpha6_r2_revised2}
\vspace{-0.4cm}
\end{figure}

\begin{figure}[htb]
\hspace{-0.3cm}
\includegraphics[scale=0.50]{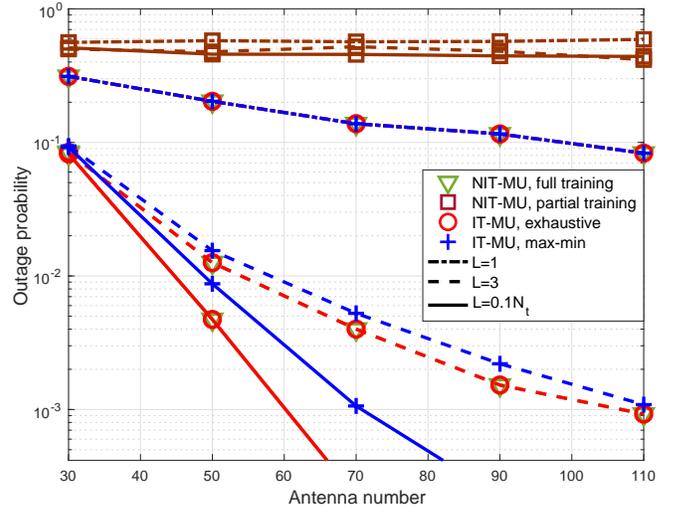}
\captionsetup{margin=5pt,font=small}
\caption{Outage performance of the IT-MU scheme for $N_{RF}=U=3$ and $\bar \alpha=6$.}\label{mu_OP_U3_alpha6_r2_revised2}
\vspace{-0.4cm}
\end{figure}

Figs.~\ref{mu_tr_U3_alpha6_r2_revised2} and \ref{mu_OP_U3_alpha6_r2_revised2} show the average training length and outage performance of the IT-MU scheme in Algorithm \ref{IMTS} respectively where $N_{RF}=U=3$
and $\bar \alpha=6$. Both the exhaustive search and the max-min method are considered for the beam assignment in the IT-MU scheme.
It can be seen that for $L=1,3$, the average training lengths of the IT-MU scheme have linear increase with $N_t$, where the slopes are approximately $0.75$ and $0.38$, respectively. Compared with the NIT-MU scheme with full training, where the training length equals $N_t$, the reduction in training length of the proposed scheme is significant, and larger $L$ results in bigger reduction.
Second, when $L=0.1N_t$, the training length of the IT-MU scheme approaches a constant, which equals $23.9$, as $N_t$ grows.
The IT-MU scheme has the same outage performance as that of the NIT-MU scheme with full training which is much better than that of the NIT-MU scheme with the same training length (partial training). Lastly, by replacing the exhaustive search with the max-min method for beam assignment, the outage performance of the IT-MU scheme has some small degradation for $L=3$ and the degradation diminishes for $L=1$. When $L=0.1N_t$, the outage performance degradation due to the sub-optimal beam assignment is more visible. On the other hand, the increment of average training length due to the use of this sub-optimal assignment method is negligible. Considering the lower complexity of the max-min method, its application in the IT-MU scheme is more desirable.

\section{Conclusions}
For the hybrid massive antenna systems, we studied the beam-based training and joint beamforming designs for SU and MU transmissions with outage probability as the performance measure. For SU systems, via concatenating the feedback with the training, an interleaved training scheme was proposed whose training length is adaptive to channel realizations. Then, exact analytical expressions were derived for the average training length and outage probability of the proposed scheme. For MU systems, we proposed a joint interleaved training and transmission design, which contains two new techniques compared to the single-user case: having the BS control the training process due to the limited local CSI at the users and feasible beam assignment methods.
Analytical and simulated results show that the proposed training and joint transmission designs achieve the same performance as the traditional full-training scheme while save the training overhead significantly. Meanwhile, useful insights were obtained on the training length and outage probability of typical network scenarios and on the effect of important system parameters, e.g., the BS antenna number, the RF chain number, the channel path number or angle spread, and the rate requirement. 

Many future directions can be envisioned on interleaved training designs for hybrid massive MIMO systems. One possible extension is to consider systems with channel estimation error and analyze how it affects the system performance. Another practical issue is the feedback overhead. For both  single-user and multi-user systems with limited feedback capacity, an important topic is the joint design of interleaved training, feedback, and transmission scheme. Interleaved training designs and analysis for multi-cell and cooperative systems are also meaningful future directions.

\vspace{-0.15cm}
\appendices
\section{The Proof of Theorem \ref{theorem1}}\label{proofD_theorem1}
In calculating the average training length, the probability that the training length is $i$ (denoted as $P_i$) for $i=1,...,N_t$ is needed. Since $\sum_{i=1}^{N_t}P_i=1$, it is sufficient to calculate $P_i, i=1,...N_t-1$ only.

The training length is $1$ when the 1st beam is a non-zero beam and its effective channel gain is strong enough to avoid outage. The probability that the 1st beam is non-zero is $L/N_t$ and 
\[{\rm Pr}(\|{\bar h}_{1}\|^2>\alpha/N_t)=\int_{{\alpha}/{N_t}}^{\infty}Le^{-Lx}dx=e^{{-\alpha L}/{N_t}}.\]
Thus \eqref{eq8} is obtained for $P_1$.

The training length is $i$ for $i \in [2, N_t-1]$ when 1) the $i$th beam is a non-zero beam, 2) an outage cannot be avoided by previously trained beams, and 3) an outage can be avoided with the newly discovered $i$th beam. To help the presentation, denote this event as Event $X$. It can be partitioned into the sub-events with respect to different $j$ for $\max\{0,L-N_t-1+i\}\le j\le  \min\{L-1, i-1\}$, where for the $j$th event, there are $j$ non-zero beams within the first $i-1$ beams and $L-1-j$ non-zero beams within the beams from $i+1$ to $N_t$. The probability for the beam distribution of sub-event $j$ thus equals to $\xi(i,j)$ as defined in (\ref{xi_ref}). Notice that  $b_i^{(1)}$ and $b_i^{(2)}$ are defined as the bounds of $j$. Further, given the the beam distribution of sub-event $j$, the probability of Event $X$ can be calculated by considering three cases as follows. To help the proof, denote the indices of the already trained $j$ non-zero beams as $n_1,\cdots,n_j$ and let $z\triangleq \|{\bar h}_{i}\|^2$.

Case 1 is when $j=0$. Event $X$ happens when $\|{\bar h}_{i}\|^2> \alpha/N_t$, whose probability is $e^{{-\alpha L}/{N_t}}$.
Case 2 is when $0<j\le N_{RF}-1$.  Event $X$ happens when $x\triangleq\sum_{l=1}^{j}\|{\bar h}_{n_l}\|^2\le{\alpha}/{N_t}$ and $z>{\alpha}/{N_t}-x$. Since $x\sim 1/(2L){\mathcal X}^2(2j)$ and $z\sim 1/(2L){\mathcal X}^2(2)$,
\begin{eqnarray}
\nonumber&&{\rm Pr}[X]={\rm Pr}\left(x\le {\alpha}/{N_t}, z>{\alpha}/{N_t}-x\right)\\
\nonumber&&\hspace{1.cm}=\int_0^{\frac{\alpha }{{{N_t}}}} {L^jx^{j-1}\frac{e^{-Lx}}{(j-1)!}\left( {\int_{\frac{\alpha }{{{N_t}}} - x}^\infty  {L{e^{ - Lz}}} dz} \right)dx}\\
\nonumber&&\hspace{1.cm}= \frac{{{{(\frac{{L\alpha }}{{{N_t}}})}^j}{e^{\frac{{ - L\alpha }}{{{N_t}}}}}}}{{j!}}.
\end{eqnarray}

Case 3 is when $j>N_{RF}-1$, where two sub-cases are considered. Case 3.1: If $N_{RF}=1$, Event $X$ happens when  $x_l\triangleq\|{\bar h}_{n_l}\|^2\le {\alpha}/{N_t}$ for all $l=1,...,j$ and $z>{\alpha}/{N_t}$. Since $x_l$'s and $z$ are i.i.d.~following ${1/(2L)\mathcal X}^2(2)$, we have ${\rm Pr}[X]={e^{\frac{{ - \alpha L}}{{{N_t}}}}}{\left( {1 - {e^{\frac{{ - \alpha L}}{{{N_t}}}}}} \right)^j}$.
Case 3.2: If $N_{RF}>1$, order the already trained $j$ non-zero beams such that $\|{\bar h}_{s_1}\|^2\ge\cdots\ge \|{\bar h}_{s_j}\|^2$, where $s_1,\cdots,s_j\in\{n_1,\cdots,n_j\}$. Event $X$ happens when $x'\triangleq\sum_{l=1}^{N_{RF}-1}\|{\bar h}_{s_{l}}\|^2\le{\alpha}/{N_t}$, $y=\|{\bar h}_{s_{{N_{RF}}}}\|^2 \le {\alpha}/{N_t}-x'$, and $z>{\alpha}/{N_t}-x'$. Notice that $x'$ and $y$ are correlated but both are independent to $z$. Via utilizing the result of the joint
distributions of partial sums of order statistics \cite[Eq. (3.31)]{Yang_order}, the joint probability density function (PDF) of $x'$ and $y$ can be given as
\begin{eqnarray}
\nonumber \hspace{-0.2cm}p(x',y)\hspace{-0.3cm}&=&\hspace{-0.4cm}\sum\limits_{l=0}^{j-N_{RF}}\beta_{j,l} [x'-(N_{RF}-1)y]^{(N_{RF}-2)}e^{-\frac{x'+(l+1)y}{1/L}},\\
\nonumber && \hspace{2.5cm}y\ge 0, x'\ge(N_{RF}-1)y.
\end{eqnarray}
Consequently, for Case 3.2,
\begin{eqnarray}\nonumber
&&\hspace{-0.8cm}{\rm Pr}[X]={\rm Pr}\left(x'\le{\alpha}/{N_t}, y \le {\alpha}/{N_t}-x', z>{\alpha}/{N_t}-x' \right)\\
\nonumber&&\hspace{0.2cm}=\hspace{-1.5mm}\int_{0}^{\hspace{-0.5mm}\frac{\alpha}{N_t}}\hspace{-.3cm}\int_{0}^{\min\left(\frac{x'}{\hspace{-1mm}N_{\hspace{-0.5mm}RF}-1},\frac{\alpha}{N_t}\hspace{-0.3mm}-\hspace{-0.3mm}x'\hspace{-0.5mm}\right)}
\hspace{-.3cm}\int_{\hspace{-0.5mm}\frac{\alpha}{N_t}-x'}^{\infty}\hspace{-0.3cm}p(x',y)Le^{-Lz}dx'dydz \\
&&\hspace{0.2cm}=P_j^{(1)}\hspace{-0.1cm}+\hspace{-0.1cm}P_j^{(2)}, \nonumber
\end{eqnarray}
where $P_j^{(1)}$ is the integral for $x'\in[0,{\frac{\alpha }{{{N_t}}}\frac{{{N_{RF}} - 1}}{{{N_{RF}}}}}]$, where $\min(\frac{x'}{N_{RF}-1},\frac{\alpha}{N_t}-x')=\frac{x'}{N_{RF}-1}$ and $P_j^{(2)}$ is that for $x'\in[{\frac{\alpha }{{{N_t}}}\frac{{{N_{RF}} - 1}}{{{N_{RF}}}}},\frac{\alpha}{N_t}]$, where $\min(\frac{x'}{N_{RF}-1},\frac{\alpha}{N_t}-x')=\frac{\alpha}{N_t}-x'$.
Via utilizing $(a+b)^{n}=\sum_{m = 0}^{n} \binom{n}{m} a^mb^{n- m}$,
the definition of the lower and upper incomplete gamma functions, and the indefinite integral  $\int x^{b-1}\Upsilon(s,x)dx=\frac{1}{b}\left[x^b\Upsilon(s,x)+\Gamma(s+b,x)\right]$, $P_j^{(1)}$ and $P_j^{(2)}$ can be derived as \eqref{P_1} and \eqref{P_2}, respectively.

Via the law of total probability and after some simple reorganizations based on the previous derivations, $P_i, i\in[2,N_t-1]$ in \eqref{eq_9} can be obtained, which completes the proof.

\section{The Proof for Lemma \ref{corollary4}}\label{appendix F}
When $N_t\gg1,L=\mathcal{O}(1)$, for the special cases of $N_{RF}=1$ or $L$, the $P_i$ values in (\ref{eq8}) and (\ref{eq_9}) can be simplified via long but straightforward calculations to the following
\[P_i=\left\{ \begin{array}{ll} \mathcal{O}(N_t^{-2}) & i> N_t+1-L \\
\frac{\binom{N_t-i}{L-1}}{{\binom{N_t}{L}}}\left[1+\mathcal{O}(N_t^{-1})\right]+\mathcal{O}(N_t^{-2}) & i\le N_t+1-L
\end{array} ,\right.\]
for $i=1,2,..., N_t-1$.
Since ${\binom{N_t-i}{L-1}}/{{\binom{N_t}{L}}}$ has the same order as or a lower order than $\mathcal{O}(N_t^{-1}), \forall i$, and
$\sum_{i=1}^{N_t-1}(N_t-i)\mathcal{O}(N_t^{-2})=\mathcal{O}(1)$, from \eqref{tr-len} we have
\begin{eqnarray}\label{eq22}
T_{\text{IT-SU}}=N_t-\sum\nolimits_{i=1}^{N_t+1-L}x_i+\mathcal{O}(1),
\end{eqnarray}
where $x_i\triangleq (N_t-i)\binom{N_t-i}{L-1}/\binom{N_t}{L}$.
We rewrite $x_i$ as
\begin{eqnarray}
\nonumber x_i\hspace{-2mm}&=&\hspace{-2mm}\frac{L(N_t-i)(N_t-i)
\times...\times(N_t-i-L+2)}{N_t\times...\times(N_t-L+1)}\\
\nonumber&=&\hspace{-2mm}\frac{L\sum_{k=0}^{L}\sum_{n=0}^{L-k}C^{(0)}_{k,n} N_t^k i^{n}}{N_t\times...\times(N_t-L+1)},
\end{eqnarray}
where $C^{(0)}_{k,n}$ is the polynomial coefficient for the term $N_t^k i^{n}$.

Define $\Delta^{(m)}_i\triangleq\Delta^{(m-1)}_{i+1}-\Delta^{(m-1)}_{i}$ for $m=1,\cdots,N_t-L$ where $\Delta^{(0)}_{i}=x_i$. Using the binomial formula, we have
$\Delta^{(m)}_i=N(\Delta^{(m)}_i)/[N_t\hspace{-1mm}\times\hspace{-1mm}...\hspace{-1mm}\times\hspace{-1mm}(N_t-L+1)]$,
where
\begin{eqnarray}
&&N(\Delta^{(m)}_i)\triangleq L\sum_{k=0}^{L}\sum_{n=0}^{L-k}C^{(0)}_{k,n} N_t^k \sum_{i_1=1}^{n}\binom{n}{i_1}\sum_{i_2=1}^{n-i_1}\binom{n-i_1}{i_2}\nonumber \\
\nonumber&&\hspace{2cm}\cdots\sum_{i_m=1}^{n-\sum_{j=1}^{m-1}i_j}\hspace{-0.2cm}\binom{n-\sum_{j=1}^{m-1}i_j}{i_m}i^{n-\sum_{j=1}^{m}i_j}.
\end{eqnarray}
Since $n\le L-k$ and $i_j\ge 1,j=1,...,m$
, we have $C^{(0)}_{k,n}=0$ for $k>L-m$, i.e., $n<m$. Thus, the highest power of $N_t$ in $N(\Delta^{(m)}_i)$ is $L-m$ and its scalar coefficient is $LC^{(0)}_{L-m,m}m!$. And this term corresponds to $i_1=...=i_m=1$, which guarantees $C^{(0)}_{L-m,n}\ne0$.
Further,
$\Delta^{(L-1)}_i$ is an arithmetic progression.
Then we have
\begin{eqnarray}
\nonumber&&\hspace{-0.8cm}\sum\nolimits_{i=1}^{N_t-L+1}x_i=\sum\nolimits_{i=1}^{N_t-L+1}x_1+
\sum\nolimits_{i=1}^{N_t-L+1}\sum\nolimits_{j_1=1}^{i-1}\Delta^{(1)}_1\hspace{-1mm}+\hspace{-1mm}...\\
\nonumber&&\hspace{-0.8cm}+\sum\nolimits_{i=1}^{N_t-L+1}\sum\nolimits_{j_1=1}^{i-1}
...\sum\nolimits_{j_{L-1}=1}^{j_{L-2}-1}(\Delta^{(L-1)}_1+\sum\nolimits_{j_{L}=1}^{j_{L-1}-1}\Delta^{(L)}_1).
\end{eqnarray}
From the Faulhaber's formula,
\[\sum\nolimits_{k=1}^{n}k^p=\frac{1}{p+1}\sum\nolimits_{j=0}^{p}\binom{p+1}{j}B_jn^{p+1-j},\]
where $B_j$ is the Bernoulli number, we have
\begin{eqnarray}
\nonumber\sum_{i=1}^{N_t-L+1}\sum_{j_1=1}^{i-1}...\sum_{j_{m}=1}^{j_{m-1}-1}\Delta^{(m)}_1
\hspace{-.2cm}&=&\hspace{-.2cm}\frac{\Delta^{(m)}_1}{({m+1})!}[N_t^{m+1}+\mathcal{O}(N_t^{m})]
\end{eqnarray}
for $m\in[1,L]$ with $j_{0}=i$.
Since the denominator of $x_1$ is $N_t^{L}+\mathcal{O}(N_t^{L-1})$ and the numerator of $x_1$ is $LN_t^{L}+ \mathcal{O}(N_t^{L-1})$, we have $(N_t-L+1)x_1=LN_t+\mathcal{O}(1)$. Consequently,
\begin{eqnarray}
\nonumber&&\hspace{-0.4cm}\sum_{i=1}^{N_t-L+1}\hspace{-2mm}x_i=LN_t\hspace{-1mm}+\hspace{-1mm}\mathcal{O}(1)\hspace{-1mm}+\hspace{-1mm}\sum_{m=1}^{L}\frac{\Delta^{(m)}_1}{({m+1})!}[N_t^{m+1}+\mathcal{O}(N_t^{m})]\\
\nonumber&&\hspace{-0.4cm}=LN_t+\mathcal{O}(1)\\
\nonumber&&\hspace{-0.4cm}+\sum_{m=1}^{L}\hspace{-1mm}
\frac{LC^{(0)}_{L-m,m}m!N_t^{L-m}\hspace{-1mm}+\hspace{-1mm}\mathcal{O}(N_t^{L-m-1})}
{[N_t^{L}+\mathcal{O}(N_t^{L-1})]({m+1})!}[N_t^{m+1}+\mathcal{O}(N_t^{m})]\\
\nonumber&&\hspace{-0.4cm}=LN_t+LN_t\sum\nolimits_{m=1}^{L}\frac{C^{(0)}_{L-m,m}}{({m+1})}+\mathcal{O}(1)\\
\nonumber&&\hspace{-0.4cm}=LN_t+LN_t \hspace{-1mm}\sum\nolimits_{m=1}^{L}\hspace{-3mm}
\frac{\binom{L}{m}(-1)^m}{{m+1}}+\mathcal{O}(1)\mathop {\rm{ = }}\limits^{(a)} \frac{L}{L+1}N_t+\mathcal{O}(1),
\end{eqnarray}
where (a) follows from $\sum_{m=1}^{L}{\binom{L}{m}(-1)^m}/{({m+1})}=-\frac{L}{L+1}$. From this result and (\ref{eq22}), \eqref{eq16} can be easily obtained.

\section{The Proof of Theorem \ref{outage_theorem}}\label{appA_Lemma1}
For a given channel realization with channel path indices ${\mathcal I}=\{I_1,...,I_L\}$, with the IT-SU scheme, an outage happens only when all $N_t$ beams have been trained and the strongest $N_{RF}$ beams among them cannot avoid an outage.
Let $\mathcal{S}_{N_t}$ be the set containing the indices of the $N_{RF}$ beams with the strongest effective channel gains. From the results on the partial sum of order statistics \cite[Eq. 3.19]{Yang_order}, the PDF of $x \triangleq \sum_{l\in \mathcal{S}_{N_t}}\|{\bar h}_{l}\|^2$ is
\vspace{-0.2cm}
\begin{eqnarray}\label{eq_23}\nonumber
&&p(x)=\frac{L!}{(L-{N_{RF}})!N_{RF}!}e^{{-Lx}}{\bigg[}\frac{L^{N_{RF}}x^{N_{RF}-1}}{(N_{RF}-1)!}\\
\nonumber&&+L\hspace{-1mm}\sum_{l=1}^{L-N_{RF}}\hspace{-2mm}
(-1)^{N_{RF}+l-1}\frac{(L-N_{RF})!}{(L-N_{RF}-l)!l!}
\left(\hspace{-1mm}\frac{N_{RF}}{l}\hspace{-1mm}\right)^{\hspace{-1mm}N_{RF}-1}\\
&&\hspace{2cm}\times\left(e^{-\frac{lxL}{N_{RF}}}-A(l,x)\right){\bigg]}, x\ge0,
\end{eqnarray}
\vspace{-0.3cm}where
\vspace{-0.1cm}
\begin{equation}\nonumber
A(l,x) \triangleq \left\{ {\begin{array}{*{20}{c}}
{\sum_{m=0}^{N_{RF}-2}\frac{1}{m!}\left(-\frac{lxL}{N_{RF}}\right)^m}, & N_{RF}\ge 2\\
0 & \text{otherwise}
\end{array}.} \right.
\end{equation}
\vspace{-0.3cm}Thus
\[{\rm out}(\text{IT-SU})={\rm Pr}(x\le\alpha/N_t),\]
which leads to \eqref{out_ITsu1} by using \eqref{eq_23}.


\section{The Proof of Lemma \ref{Corollary 1}}\label{appendB_Corollary1}
Since
$\Upsilon \hspace{-1mm}\left( {1,\hspace{-1mm}\frac{{L\alpha }}{{{N_t} }}} \right)\hspace{-1mm}=\hspace{-1mm}1-e^{-\frac{L \alpha}{N_t}}$ and $N_{RF}\hspace{-1mm}=\hspace{-1mm}1$,
from \eqref{out_ITsu1}, we have
\vspace{-0.3cm}
\begin{eqnarray}
\nonumber&&\hspace{-0.5cm}{\rm out}({\text{IT-SU}})=L\sum\nolimits_{l=0}^{L-1}(-1)^l\frac{(L-1)!}{(L-1-l)!l!}\frac{e^{(-1-l)\frac{L \alpha }{N_t}}-1}{-1-l}\\
\nonumber&&\hspace{-0.5cm}=\sum\nolimits_{l=0}^{L-1}(-1)^{l+1}\frac{L!}{(L-(l+1))!(l+1)!}\left({e^{-(1+l)\frac{L \alpha}{N_t}}}-1\right)\\
\nonumber&&\hspace{-0.5cm}\mathop {\rm{ = }}\limits^{(a)}(-1)^L\sum\nolimits_{t=0}^{L}(-1)^{L-t}\frac{L!}{(L-t)!t!}e^{-t\frac{L \alpha}{N_t}}-1\\
\nonumber&&\hspace{-0.5cm}-\sum\nolimits_{t=0}^{L}(-1)^t\frac{L!}{(L-t)!t!}+1\mathop {\rm{ = }}\limits^{(b)}(1-e^{-\frac{L\alpha}{N_t}})^L,
\end{eqnarray}
where (a) and (b) follow from the variable substitutions  $t=l+1$ and  $(x+y)^n=\sum_{l=0}^{n}\binom{n}{l}x^{n-l}y^l$, respectively.


\ifCLASSOPTIONcaptionsoff
  \newpage
\fi

\bibliographystyle{IEEEtran}

\begin{IEEEbiography}[{\includegraphics[width=1in,height=1.25in,clip,keepaspectratio]{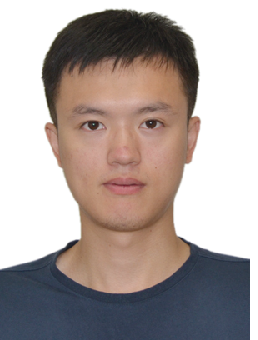}}]{Cheng Zhang} received the B.Eng. degree from Sichuan University, Chengdu, China in June 2009, and the M.Sc. degree from Institute No.206 of China Arms Industry Group Corporation, Xian, China in May 2012. He worked as a Radar signal processing engineer at Institute No.206 of China Arms Industry Group Corporation, Xian, China from June 2012 to Aug. 2013. Since Mar. 2014, he has been working towards Ph.D. degree at Southeast University, Nanjing, China. From Nov. 2016 to Nov. 2017, he was a visiting student with University of Alberta, Edmonton, Canada. His current research interests include space-time signal processing and application of learning algorithm in channel estimation and transmission design for millimeter-wave massive MIMO communication systems.
\end{IEEEbiography}

\begin{IEEEbiography}[{\includegraphics[width=1in,height=1.25in,clip,keepaspectratio]{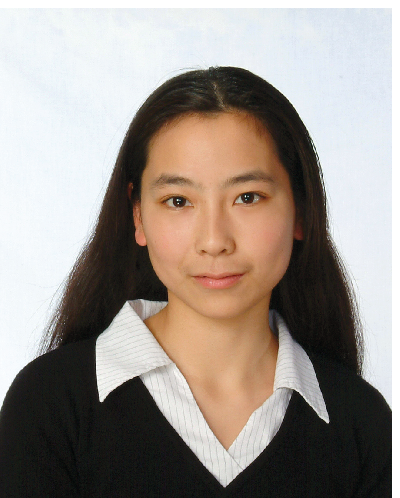}}]{Yindi Jing} received the B.Eng. and M.Eng. degrees from the University of Science and Technology of China, in 1996 and 1999, respectively. She received the M.Sc. degree and the Ph.D. in electrical engineering from California Institute of Technology, Pasadena, CA, in 2000 and 2004, respectively. From Oct. 2004 to Aug. 2005, she was a postdoctoral scholar at the Department of Electrical Engineering of California Institute of Technology. Since Feb. 2006 to Jun. 2008, she was a postdoctoral scholar at the Department of Electrical Engineering and Computer Science of the University of California, Irvine. In 2008, she joined the Electrical and Computer Engineering Department of the University of Alberta, where she is currently an associate professor. She was an Associate Editor for the IEEE Transactions on Wireless Communications 2011-2016 and currently serves as a Senior Area Editor for IEEE Signal Processing Letters (since Oct. 2017) and a member of the IEEE Signal Processing Society Signal Processing for Communications and Networking (SPCOM) Technical Committee. Her research interests are in massive MIMO systems, cooperative relay networks, training and channel estimation, robust detection, and fault detection in power systems.
\end{IEEEbiography}

\begin{IEEEbiography}[{\includegraphics[width=1in,height=1.25in,clip,keepaspectratio]{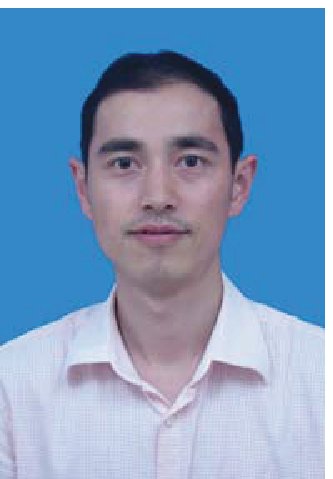}}]{Yongming Huang} received the B.S. and M.S. degrees from Nanjing University, Nanjing, China, in 2000 and 2003, respectively, and the Ph.D. degree in electrical engineering from Southeast University, Nanjing, China, in 2007. Since 2007, he has been a Faculty Member with the School of Information Science and Engineering, Southeast University, where he is currently a Full Professor. From 2008 to 2009, he visited the Signal Processing Laboratory, School of Electrical Engineering, Royal Institute of Technology, Stockholm, Sweden. He has authored over 200 peer-reviewed papers, hold over 50 invention patents, and submitted over 10 technical contributions to the IEEE standards. His current research interests include MIMO wireless communications, cooperative wireless communications, and millimeter wave wireless communications. He has served as an Associate Editor for the IEEE TRANSACTIONS ON SIGNAL PROCESSING, IEEE WIRELESS COMMUNICATIONS LETTERS, EURASIP Journal on Advances in Signal Processing, and EURASIP Journal on Wireless Communications and Networking.

\end{IEEEbiography}
\begin{IEEEbiography}[{\includegraphics[width=1in,height=1.25in,clip,keepaspectratio]{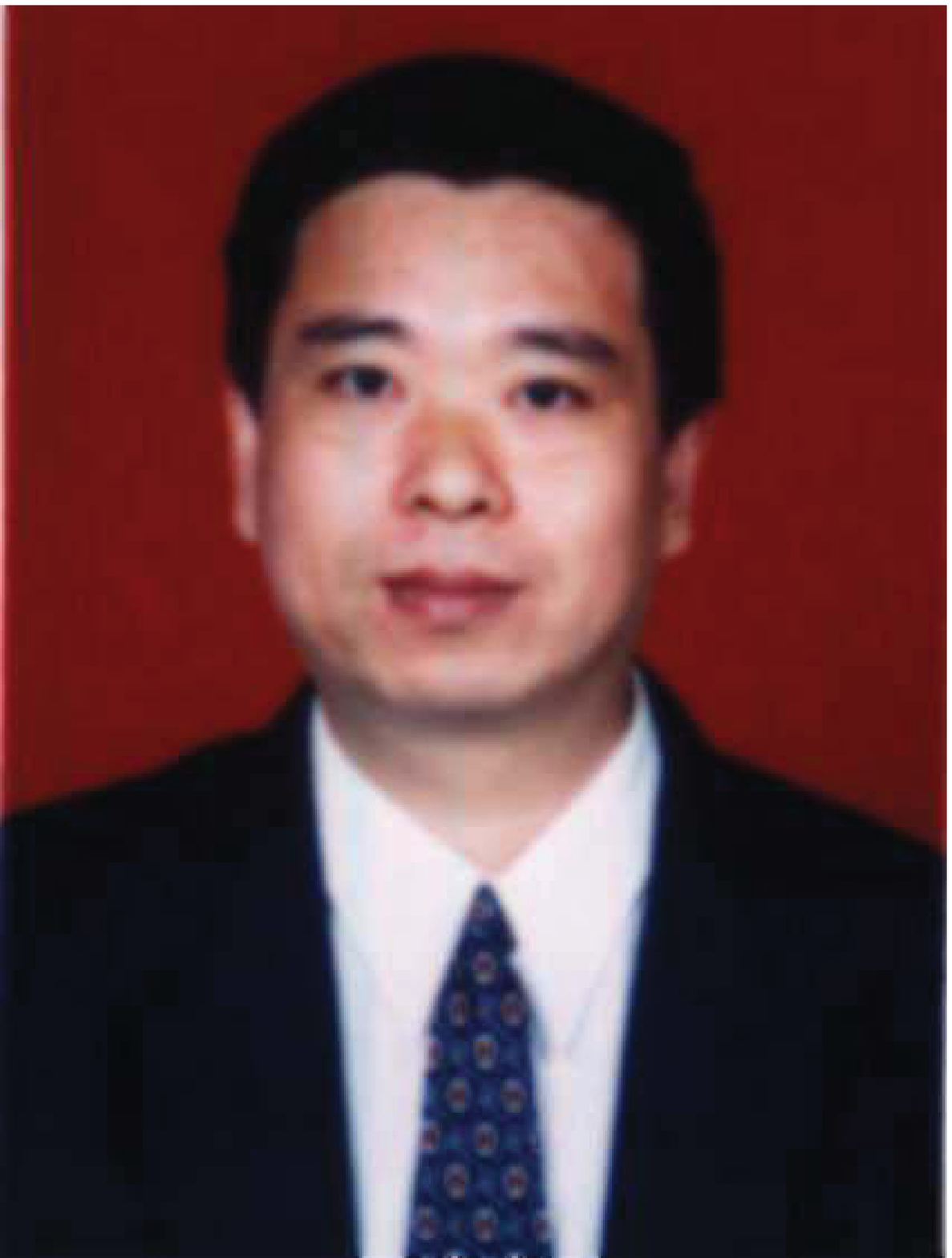}}]{Luxi Yang} received the M.S. and Ph.D. degree in electrical engineering from the Southeast University, Nanjing, China, in 1990 and 1993, respectively. Since 1993, he has been with the School of Information Science and Engineering, Southeast University, where he is currently a full professor of information systems and communications, and the Director of Digital Signal Processing Division. His current research interests include signal processing for wireless communications, MIMO communications, cooperative relaying systems, and statistical signal processing. He has authored or co-authored two published books and more than 150 journal papers, and holds 30 patents. Prof. Yang received the first- and second-class prizes of Science and Technology Progress Awards of the State Education Ministry of China in 1998, 2002 and 2014. He is currently a member of Signal Processing Committee of Chinese Institute of Electronics.
\end{IEEEbiography}

\end{document}